\documentclass[12pt]{article}
\pdfoutput=1
\usepackage{CustomTitle}
\usepackage[margin=1in]{geometry}

\usepackage{amsmath}
\usepackage{amssymb}
\usepackage{graphicx}
\usepackage[bf,margin=20pt,font={small}]{caption}
\usepackage{color}
\usepackage{colortbl}
\makeatletter\let\over\@@over\makeatother    

\usepackage{tikz}
\usetikzlibrary{arrows}

\usepackage[colorlinks=true, urlcolor=dblue,linkcolor=dblue, citecolor=purple]{hyperref}
\usepackage{xcolor}
\colorlet{dblue}{blue!70!black}

\newcommand\padic{$p$-adic }

\newcommand\Ok{{\mathcal{O}_K^\times}}

\newcommand\be{\begin{equation}}
\newcommand\ba{\begin{eqnarray}}
\newcommand\ee{\end{equation}}
\newcommand\ea{\end{eqnarray}}

\definecolor{Gray}{gray}{0.9}
\newcommand\TL{\hfil$\displaystyle{##}$}
\newcommand\TR{$\displaystyle{{}##}$\hfil}

\newcommand\TT{\hbox{##}}
\def\seqalign#1#2{\vcenter{\openup1\jot
  \halign{\strut #1\cr #2 \cr}}}
\def\lbldef#1#2{\expandafter\gdef\csname #1\endcsname {#2}}
\newcommand{\eqn}[3][]{\lbldef{#2}{(\ref{#2})}%
\begin{equation} \eqalign{#3} \label{#2} \end{equation}}
\def\eqalign#1{\vcenter{\openup1\jot
    \halign{\strut\span\TL & \span\TR\cr #1 \cr
   }}}
\def\eno#1{(\ref{#1})}
\def\href#1#2{#2}
\def\mop#1{\mathop{\rm #1}\nolimits}
\def\sgn{\mop{sgn}}
\def\mod{\mop{mod}}
\newcommand{\cir}[1]{\textcircled{\raisebox{.8pt}{\tiny #1}}}
\begin{document}

\preprint{BRX-TH 6321, CALT-TH 2017-036, PUPT-2527}
\title{Signs of the time:\\[-6pt] Melonic theories over diverse number systems}
\authors{Steven S. Gubser,$^\Princeton$ Matthew Heydeman,$^\Burke$ Christian Jepsen,$^\Princeton$\\[3pt] Sarthak Parikh,$^\Princeton$ Ingmar Saberi,$^\Heidelberg$ Bogdan Stoica,$^{\Brandeis,\Brown}$ and Brian Trundy$^\Princeton$\\[10pt] {\tt\footnotesize ssgubser@princeton.edu, mheydema@caltech.edu, cjepsen@princeton.edu, sparikh@princeton.edu, saberi@mathi.uni-heidelberg.de, bstoica@brandeis.edu, btrundy@princeton.edu}
}

\institution{Princeton}{$^\Princeton$Joseph Henry Laboratories, Princeton University, Princeton, NJ 08544, USA}
\institution{Burke}{$^\Burke$Walter Burke Institute for Theoretical Physics,\cr\hskip0.06in California Institute of Technology, 452-48, Pasadena, CA 91125, USA}
\institution{Heidelberg}{$^\Heidelberg$Mathematisches Institut, Ruprecht-Karls-Universit\"at Heidelberg, 
\cr\hskip0.06in Im Neuenheimer Feld 205, 69120 Heidelberg, Germany}
\institution{Brandeis}{$^\Brandeis$Martin A. Fisher School of Physics, Brandeis University, Waltham, MA 02453, USA}
\institution{Brown}{$^\Brown$Department of Physics, Brown University, Providence RI 02912, USA}
\date{July 2017}
\abstract{Melonic field theories are defined over the $p$-adic numbers with the help of a sign character. Our construction works over the reals as well as the $p$-adics, and it includes the fermionic and bosonic Klebanov-Tarnopolsky models as special cases; depending on the sign character, the symmetry group of the field theory can be either orthogonal or symplectic. Analysis of the Schwinger-Dyson equation for the two-point function in the leading melonic limit shows that power law scaling behavior in the infrared arises for fermionic theories when the sign character is non-trivial, and for bosonic theories when the sign character is trivial. In certain cases, the Schwinger-Dyson equation can be solved exactly using a quartic polynomial equation, and the solution interpolates between the ultraviolet scaling controlled by the spectral parameter and the universal infrared scaling. As a by-product of our analysis, we see that melonic field theories defined over the real numbers can be modified by replacing the time derivative by a bilocal kinetic term with a continuously variable spectral parameter. The infrared scaling of the resulting two-point function is universal, independent of the spectral parameter of the ultraviolet theory.}

\maketitle

\section{Introduction}

A first working definition of a $p$-adic quantum field theory is a theory defined through a functional integral over maps $\phi\colon \mathbb{Q}_p \to \mathbb{R}$, where $\mathbb{Q}_p$ denotes the $p$-adic numbers and $\mathbb{R}$ denotes the reals.\footnote{In physics parlance, $\phi$ is termed the field; unfortunately, in math parlance, $\mathbb{Q}_p$ and $\mathbb{R}$ are fields.  In this paper we will overload the term ``field'' to carry both definitions, on the expectation that context will make the meaning clear.}  We may expand our definition of $p$-adic quantum field theories by replacing $\mathbb{Q}_p$ with a field extension of $\mathbb{Q}_p$, and by allowing $\phi$ to be valued in some vector space over $\mathbb{R}$.  If the values of $\phi$ are anti-commuting, we refer to $\phi$ as a fermionic field, while if they are commuting, we refer to $\phi$ as a bosonic field.

\begin{figure}[h]\centerline{\includegraphics{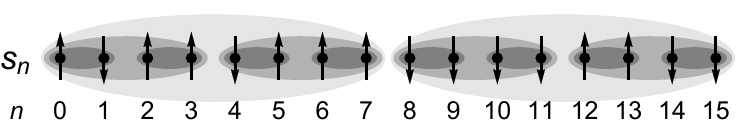}}
  \caption{The hierarchical model, with an Ising spin $s_n = \pm 1$ at every integer point $n$, and successively weaker ferromagnetic couplings between pairs of spins, pairs of pairs, and so on.}\label{Hierarchical}
 \end{figure}

A first inkling of $p$-adic field theories (for $p=2$) appeared in the form of the Dyson hierarchical model \cite{Dyson:1968up}.  In this model, one starts with a chain of Ising spins $s_n$ where $n \in \mathbb{Z}$.  A strong ferromagnetic interaction is assigned between spin $2n$ and spin $2n+1$ for all $n$.  Next one assigns a weaker ferromagnetic interaction between pairs, and yet a weaker ferromagnetic interaction between pairs of pairs.  After $\ell-1$ steps, one has blocks of $2^{\ell-1}$ spins, and in the $\ell$-th step one pairs up neighboring blocks with a ferromagnetic interaction whose strength decreases as a power of $2^\ell$: See figure~\ref{Hierarchical}.
The critical behavior of this model is described by a $2$-adic conformal field theory, meaning a $2$-adic field theory which is invariant under the action of ${\rm PGL}(2,\mathbb{Q}_2)$ on $\mathbb{Q}_2$ through linear fractional transformations, $t \to {at+b \over ct+d}$, where $a,b,c,d$ as well as $t$ are in $\mathbb{Q}_2$.  We may approximately understand the action of these linear fractional transformations as mapping spin blocks to spin blocks, but not necessarily preserving the size of the blocks.  In practical terms, ${\rm PGL}(2,\mathbb{Q}_p)$ invariance restricts the form of correlators of local operators: for example, in scalar $p$-adic field theories we have $\langle \phi(t) \phi(0) \rangle = K/|t|^\Delta$ where $K$ and $\Delta$ are real constants, and $|t|$ is the $p$-adic norm.

Hierarchical constructions are conducive to Wilsonian renormalization group ideas, because the hierarchical structure provides a natural way to perform Kadanoff spin blocking \cite{Kadanoff:1966wm}.  Indeed, the early literature on the Wilsonian renormalization group shows that the original practitioners were well aware of hierarchical models, if not their connection to the $p$-adic numbers; see for example the review \cite{Wilson:1973jj}.  Subsequently, renormalization group flows in $p$-adic field theories have been studied fairly extensively, notably by Missarov and collaborators; see the recent review \cite{missarov2012p}.

Recently, progress has been made in understanding a $p$-adic version of the anti-de Sitter / conformal field theory correspondence (AdS/CFT) \cite{Gubser:2016guj,Heydeman:2016ldy}.  The $p$-adic version of the correspondence is an equivalence between a conformal field theory on $\mathbb{Q}_p$ (more properly $\mathbb{P}^1(\mathbb{Q}_p)$) and a bulk theory defined on the Bruhat-Tits tree.  The Bruhat-Tits tree is the $p$-adic version of anti-de Sitter space, and it is the quotient of the $p$-adic conformal group ${\rm PGL}(2,\mathbb{Q}_p)$ by its maximal compact subgroup.  Particularly if we stick with $\mathbb{Q}_p$ on the boundary, rather than some extension of $\mathbb{Q}_p$, $p$-adic AdS/CFT seems to be an analog of ${\rm AdS}_2/{\rm CFT}_1$.  To develop the correspondence further, we are therefore interested in $p$-adic analogs of one-dimensional conformal theories which are thought to have ${\rm AdS}_2$ duals.

An example ${\rm CFT}_1$ of recent interest is the infrared limit of the melonic theory introduced in \cite{Witten:2016iux} and simplified in \cite{Klebanov:2016xxf}.  These theories are inspired on one hand by the Sachdev-Ye-Kitaev (SYK) model \cite{Sachdev:1992fk,Kitaev:2015zz}, and on the other by melonic scaling limits; for a review of melonic scaling see \cite{Gurau:2011xp}.  For our purposes, the melonic field theories of \cite{Witten:2016iux,Klebanov:2016xxf} are attractive for three reasons.  First, the two-point function of the fermionic field in the model can be calculated explicitly in the melonic limit thanks to a Schwinger-Dyson equation that resums precisely the diagrams that survive in the leading melonic limit.  Second, there is a non-trivial renormalization group flow from a free theory in the ultraviolet to a conformal theory in the infrared.  Third, there are indications of the existence of an ${\rm AdS}_2$ dual.

In the current work, our aim is to formulate $p$-adic versions of the simplest melonic theories.  As in our previous work on the ${\rm O}(N)$ model \cite{Gubser:2017vgc}, it is possible to give a remarkably uniform presentation in which theories over the reals and over the $p$-adics are treated on almost an equal footing.  There are three main differences between the reals and the $p$-adics:
 \begin{itemize}
  \item In $p$-adic field theories, ordinary derivatives like $d\phi/dt$ are not good starting points for kinetic terms in the action, because smooth functions $\phi(t)$ are piecewise constant.  One is driven instead to the Vladimirov derivative, which leads to bilocal kinetic terms, approximately of the form $(\phi(t_1)-\phi(t_2))/|t_1-t_2|^{1+s}$ where $s \in \mathbb{R}$ is the order of the derivative.  It is therefore natural to consider at the same time bilocal kinetic terms in theories over the reals.\footnote{While this work was nearing completion, we received \cite{Gross:2017vhb}, which also considers SYK-like models with bilocal kinetic terms.  However, the focus of \cite{Gross:2017vhb} is rather different from ours; their aim is to consider theories over $\mathbb{R}$ not $\mathbb{Q}_p$, and to deform the free bilocal theory by a marginal deformation so as to eventually reach the strongly coupled infrared theory, with every intermediate theory preserving the full ${\rm SL}(2,\mathbb{R})$ symmetry characteristic of one-dimensional conformal field theories.  We instead add  a relevant deformation to the free bilocal theory, and under appropriate conditions we find that the two-point function shows universal infrared behavior.
}
  \item A key ingredient both in the action and in Green's functions of melonic theories is the sign function $\sgn t$, which ordinarily is $-1$ if $t<0$ and $+1$ if $t>0$.  The $p$-adic numbers are not naturally ordered; however, they do admit an assortment of sign functions which are multiplicative characters: $\sgn(t_1 t_2) = (\sgn t_1)(\sgn t_2)$.  We will wind up with approximately ten variants (depending on how we count) of the simplest melonic theory due to this profusion of sign characters, which can naturally be parametrized by a non-zero $p$-adic number $\tau$.  Sign characters on the reals can be parametrized the same way, and $\tau<0$ corresponds to the ordinary notion of sign while $\tau>0$ corresponds to the trivial character which maps all real numbers to $+1$.
  \item For some sign characters, in order to get a theory with well-defined scaling behavior both in the ultraviolet and in the infrared, we are obliged to alter the global symmetry group of the theory from ${\rm O}(N)^3$ to ${\rm Sp}(N)^3$, where ${\rm Sp}(N)$ is the non-compact group of real-valued $N \times N$ matrices preserving a symplectic structure.
 \end{itemize}
The structure of the rest of the paper is as follows.  In section~\ref{KINETIC} we construct the kinetic terms which specify the theories of interest in the ultraviolet.  In section~\ref{INTERACTIONS} we add in the interaction term, which is a relevant deformation.  In section~\ref{DIAGRAMS} we explain how the Schwinger-Dyson equation for the two-point function arises.  In section~\ref{MULTIPLICATIVE} we present a self-contained account of sign characters on $\mathbb{Q}_p$ and certain extensions, including generalized $\Gamma$ and ${\rm B}$ functions of these characters.  In section~\ref{SOLUTION} we show how to solve the Schwinger-Dyson equation in the infrared.  In some cases we are able to solve it exactly at all scales.  In section~\ref{WILSON} we explain the Wilsonian perspective on renormalization of melonic theories over $\mathbb{Q}_p$, which features a powerful non-renormalization theorem on the kinetic term of the Wilsonian action.  We conclude in section~\ref{SUMMARY} with a summary and an indication of some directions for future work.  In appendix~\ref{SYMPLECTIC} we discuss symplectic groups in more detail than we give in the main text.

Readers wishing to see a tabulation of our main results can consult table~\ref{ExplicitResults}, in which we indicate for each field ($\mathbb{R}$ or $\mathbb{Q}_p$) and each sign character (parametrized by $\tau$) what sort of theory we must consider in order to have a renormalization group flow from a free ultraviolet theory to the universal infrared scaling behavior of the two-point function characteristic of SYK-type theories.

\section{The kinetic term}
\label{KINETIC}

The SYK model \cite{Sachdev:1992fk,Kitaev:2015zz}, Witten's melonic version \cite{Witten:2016iux}, and the Klebanov-Tarnopolsky model \cite{Klebanov:2016xxf} are all based on interacting Majorana fermions in $0+1$ dimensions.  The kinetic term for a single Majorana fermion is\footnote{The measure in the $\omega$ integral is $d\omega$ instead of ${d\omega \over 2\pi}$ due to our convention for Fourier transforms: $\psi(t) = \int_{\mathbb{R}} d\omega \, e^{-2\pi i \omega t} \psi(t)$.  This convention is far more convenient than the usual one when one passes to $\mathbb{Q}_p$, and it tends even to simplify formulas in the real case.}
 \eqn{SingleFermion}{
  S_{\rm free} = \int_{\mathbb{R}} dt \, {i \over 2} \psi \partial_t \psi 
   = \int_{\mathbb{R}} d\omega \, {1 \over 2} \psi(-\omega) \omega \psi(\omega) \,.
 }
The associated two-point Green's function is
 \eqn{SFG}{
  F(\omega) = {i \over \omega} \,,
 }
where we use $F(\omega)$ instead of $G(\omega)$ to emphasize that this is the Green's function of the free theory.

We would like to consider $p$-adic field theories with a similar structure. 
In order to get started, we need something that replaces $i\partial_t$.  It is easiest intuitively to work in momentum space, where in the Archimedean setting the kernel of $S_{\rm free}$ is $\Gamma_2^{\rm free}(\omega) = \omega = |\omega| \sgn \omega$.  We would like to generalize this to
 \eqn{GammaFree}{
  \Gamma_2^{\rm free}(\omega) = |\omega|^s \sgn\omega \,,
 }
where the spectral parameter $s$ is real (assuming we want a real-valued action).  We can now allow $\omega$ to be valued in a field $K$, which may be either Archimedean or ultrametric. When $K=\mathbb{R}$, the norm $|\cdot |$ in \eno{GammaFree} denotes the usual absolute value norm, while $|\cdot |$ denotes the $p$-adic norm if $K=\mathbb{Q}_p$. The sign function $\sgn\omega$ is, by assumption, a multiplicative character from $K^\times$ to $\{\pm 1\}$ (just as the ordinary sign function is). Here and below, for any ring $R$ (including if $R$ is a field), $R^\times$ means all elements of $R$ which have a multiplicative inverse.  Since $K$ is a field, $K^\times$ is all non-zero elements of $K$.

For $K=\mathbb{R}$, there are but two sign functions: the usual sign function and the trivial sign function, which evaluates to one on all non-zero numbers. If one sets $K=\mathbb{R}$ and $s=2$ and picks the trivial sign character, \eno{GammaFree} gives the standard kernel for a bosonic theory. A surprising and important point is that for ultrametric $K$ not all non-trivial sign characters have $\sgn(-1) = -1$; those that do are called odd, while those that have $\sgn(-1) = 1$ are called even.  To get the idea of why non-trivial even sign characters can be defined on some $K$, consider a simpler problem: sign characters on finite fields
 \eqn{FpDef}{
  \mathbb{F}_p = {\mathbb{Z} \over p \mathbb{Z}} = \{0,1,2,\dots,p-1\} \,,
 }
where $p$ is prime, and where addition and multiplication are defined modulo $p$.  
By definition, these are homomorphisms from $\mathbb{F}_p^\times$ to $\{\pm 1\}$.  The case $p=2$ is vacuous because $\mathbb{F}_2^\times$ is the trivial group consisting of only the identity. 
For odd~$p$, there are precisely two distinct sign characters on $\mathbb{F}_p^\times$: 
the trivial one which assigns $+1$ to all elements of $\mathbb{F}_p^\times$, and the non-trivial one, usually written as the Legendre symbol $(a|p)$, which is $1$ if $a=b^2$ for some $b \in \mathbb{F}_p^\times$ and $-1$ otherwise.  The trivial sign character is obviously even, while the non-trivial character may be even or odd:
 \eqn{SignMinusOne}{
  \sgn(-1) = (-1|p) = (-1)^{(p-1)/2} \qquad\hbox{on}\qquad 
   \mathbb{F}_p^\times \,.
 }
 For example, $\sgn(-1) = -1$ if $p=3$, corresponding to the fact that $-1 = 2 \in \mathbb{F}_3$ is not a square; but $\sgn(-1) = 1$ if $p=5$ because $-1 = 4 = 2^2$ in $\mathbb{F}_5$.  It is perhaps unsurprising that similar properties carry over to $\mathbb{Q}_p$: Non-trivial sign characters exist for all $p$, and when $p \equiv 3 \mod 4$, there are both odd and even sign characters, whereas if $p \equiv 1 \mod 4$, all sign characters are even. Sign characters are reviewed and discussed more systematically later in \ref{MULTIPLICATIVE}.

Assuming $\psi$ is a real Grassmann field, we have
 \eqn{Antisymmetry}{
  S_{\rm free} &\equiv 
    \int_K d\omega \, {1 \over 2} \psi(-\omega) \Gamma_2^{\rm free}(\omega) \psi(\omega)
   = \int_K d\omega \, {1 \over 2} \psi(\omega) \Gamma_2^{\rm free}(-\omega) \psi(-\omega)
   = -\sgn(-1) S_{\rm free} \,.
 }
The integrations over $K$ use the additive Haar measure.  In the second equality of \eno{Antisymmetry} we performed the $u$-substitution $\omega \to -\omega$.  In the next step we anti-commuted $\psi(\omega)$ past $\psi(-\omega)$ to pick up the explicit sign, and we used $\Gamma_2^{\rm free}(-\omega) = \sgn(-1) \Gamma_2^{\rm free}(\omega)$.  Thus the action we have proposed makes sense only when the sign character is odd~\cite{Marshakov:1989jz}.  We can however consider two straightforward generalizations: adding index structure to the field, and changing its statistics.  

Consider then the action
 \eqn{KTO}{
  S_{\rm free} \equiv \int_K d\omega \, {1 \over 2} \psi^i(-\omega) \Gamma_{2,ij}^{\rm free}(\omega)
    \psi^j(\omega) \qquad\hbox{where}\qquad
   \Gamma_{2,ij}^{\rm free}(\omega) = \Omega_{ij} |\omega|^s \sgn\omega \,.
 }
We assume that $\psi$ is either commuting or anti-commuting, and correspondingly we write $\sigma_\psi = +1$ or $-1$.  We refer to the commuting case as bosonic and the anti-commuting case as fermionic.  Also we assume that $\psi$ is real, i.e.~$\psi^* = \psi$.  We assume that $\Omega$ is either symmetric or antisymmetric as a matrix, and correspondingly we write $\sigma_\Omega = +1$ or $-1$.  Then the same manipulation we used to obtain \eno{Antisymmetry} generalizes immediately to give
 \eqn{KTOsym}{
  S_{\rm free} = \sigma_\psi \sigma_\Omega \sgn(-1) S_{\rm free} \,.
 }
Evidently, we want the combined sign $\sigma_\psi \sigma_\Omega \sgn(-1)$ to be equal to $1$.  Similar considerations show that in order for $S_{\rm free}$ to be real, we must choose $\Omega$ to be Hermitian.  This is assuming that conjugation reverses the order of anti-commuting factors: $(\psi^i \psi^j)^* = (\psi^j)^* (\psi^i)^*$.

\section{The full model}
\label{INTERACTIONS}

For simplicity, let's focus on a generalization of the Klebanov-Tarnopolsky model rather than the SYK model or Witten's version:
 \eqn{KTfull}{
  S = S_{\rm free} + S_{\rm int} \,,
 }
where 
 \eqn{Sfree}{
  S_{\rm free} = \int_K d\omega \, {1 \over 2} \psi^{a_1 b_1 c_1}(-\omega)
    \Omega_{a_1 a_2} \Omega_{b_1 b_2} \Omega_{c_1 c_2} |\omega|^s (\sgn\omega)
     \psi^{a_2 b_2 c_2}(\omega)
 }
and
 \eqn{Sint}{
  S_{\rm int} = \int_K dt \, {g \over 4}
    \Omega_{a_1 a_2} \Omega_{a_3 a_4} 
    \Omega_{b_1 b_3} \Omega_{b_2 b_4} 
    \Omega_{c_1 c_4} \Omega_{c_2 c_3} 
    \psi^{a_1 b_1 c_1}(t) \psi^{a_2 b_2 c_2}(t) \psi^{a_3 b_3 c_3}(t) \psi^{a_4 b_4 c_4}(t) \,.
 }
Note that $K$ can be $\mathbb{R}$ or $\mathbb{Q}_p$, or some finite extension of one of these fields.  The indices $a_i$, $b_i$, and $c_i$ all take values from $1$ to $N$, so overall the model is based on exactly $N^3$ Majorana fermions, or $N^3$ real scalars.  We assume that $N$ is even, and that
 \eqn{OmegaChoice}{
  \Omega = [\Omega_{ab}] = \left\{ \seqalign{\span\TR &\qquad\span\TT}{
   {\bf 1}_{N \times N} & if $\sigma_\Omega = 1$  \cr
   \sigma_2 \otimes {\bf 1}_{{N \over 2} \times {N \over 2}} & 
   if $\sigma_\Omega = -1$\,,
   } \right.
 }
where $\sigma_2 = \tiny \begin{pmatrix} 0 & -i \\ i & 0 \end{pmatrix}$.  The field $\psi^{abc}$ has no particular symmetry under permutations of its indices.  So we see that the full model has an ${\rm O}(N)^3$ symmetry if $\sigma_\Omega = 1$, and an ${\rm Sp}(N)^3$ symmetry if $\sigma_\Omega = -1$.  Here 
 \eqn{SpNnotation}{
 {\rm Sp}(N) \equiv {\rm Sp}(N,\mathbb{R})
 }
 is the group of $N \times N$ matrices $\Lambda^a{}_b$ with real entries such that $\Lambda^a{}_{c} \Lambda^b{}_d \Omega_{ab} = \Omega_{c d}$.  Note that ${\rm Sp}(N)$ is a {\it non-compact} group, unlike ${\rm O}(N) = {\rm O}(N,\mathbb{R})$, but at least at the level of a global symmetry group this is not a problem. See appendix~\ref{SYMPLECTIC} for discussion of our notations and conventions, which may differ from other choices in the literature.

We assume that $K$ has a finite dimension $n$ over the base field.  If the base field is $\mathbb{R}$, then the only possibilities are $n=1$ and $n=2$, because for $n>2$ there is no field structure possible in $\mathbb{R}^n$.  If the base field is $\mathbb{Q}_p$, then $n$ can be any positive integer. Importantly, we follow conventions of \cite{Gelfand:1968} by defining $|\cdot|$ so that under a $u$-substitution $u = \lambda t$, the integration measure transforms as
 \eqn{usubMeasure}{
  du = |\lambda| dt \,.
 }
For instance, when $K$ is an extension of $\mathbb{Q}_p$, we have
 \eqn{KNorm}{
  |t|_K \equiv |N_{K:\mathbb{Q}_p}(t)|_{\mathbb{Q}_p} \,,
 }
where $N_{K:\mathbb{Q}_p}$ denotes the field extension norm.  This contrasts with the conventions, e.g., of \cite{Gubser:2017vgc}; in particular, $K$ admits a unique field norm that coincides with the standard norm on $\mathbb{Q}_p\subset K$, but this field norm on $K$ would be written $|\cdot|^{1/n}$ in our present notation.  Similarly, in the current work, over $\mathbb{C}$ we have $|x+iy| = x^2+y^2$ and $dz = dx dy$.

In order to develop some first expectations about the behavior of the field theory, let's do a little dimension counting.  The following assignments are equally valid for Archimedean and ultrametric places:
 \eqn{DimCounting}{\seqalign{\span\TL & \span\TR &\qquad\qquad \span\TL & \span\TR}{
  [|\omega|] &= -[|t|] = 1 &
  [d\omega] &= -[dt] = 1   \cr
  [\psi(\omega)] &= -{1+s \over 2} &
  [\psi(t)] &= {1-s \over 2}  \cr 
  [g] &= 2s - 1 \,.
 }}
In the usual $0+1$-dimensional model with $s=1$, we have $[g] = 1$, so the interaction is a relevant deformation.  Let's assume that $s > 1/2$ for our generalized model, so that the quartic interaction is still relevant.  Then we expect to find correlators agreeing with the free theory in the ultraviolet, and some new fixed point in the infrared, analogous to the infrared behavior of the SYK model and its melonic cousins.  We should also stipulate $s\leq 1$, because if instead $s>1$, then $\psi(t)$ would have negative dimension, which signals a pathology; in particular, it means that operators schematically of the form $\psi^{2r}$ are more and more relevant as $r$ becomes large, and it is no longer clear that it is meaningful to limit ourselves to a polynomial lagrangian.\footnote{The bosonic model of \cite{Klebanov:2016xxf} in $d$ dimensions has $s=2/d$, and $d=1$ is below its lower critical dimension.  Thus, when we consider the bosonic model on $\mathbb{R}$, it should be understood as having a bilocal kinetic term with $s$ between $1/2$ and $1$.\label{LowerCriticalDim}}

\section{Diagrammatics and the Schwinger-Dyson equation}
\label{DIAGRAMS}

The diagrammatic structure of the theory \eno{KTfull} is similar to \cite{Witten:2016iux,Klebanov:2016xxf}, with the main difference being the addition of factors of $\Omega_{ab}$ and $\Omega^{ab}$, where the latter is defined by the condition
 \eqn{OmegaInverse}{
  \Omega^{ab} \Omega_{bc} \equiv \delta^a_c \,.
 }
Of course, we also have $\Omega^{ab} \Omega_{cb} = \sigma_\Omega \delta^a_c$ because of the symmetry or antisymmetry of $\Omega$.  The free propagator following from \eno{Sfree} is
 \eqn{FreeProp}{
  F^{a_2b_2c_2,a_1b_1c_1}(t) = \Omega^{a_2a_1} \Omega^{b_2b_1} \Omega^{c_2c_1} F(t) \,,
 }
where
 \eqn{FwForm}{
  F(\omega) = \sqrt{\sgn(-1)}\,{\sgn\omega \over |\omega|^s} \,.
 }
The pre-factor of $\sqrt{\sgn(-1)}$ ensures that the free propagator is real in position space, which is required for a real action in Euclidean space, as discussed at the end of this section. 
For the kinetic term in \eno{KTO} the appropriate square root to pick is
 \eqn{sqrtSgn}{
  \sqrt{\sgn(-1)} = \left\{ \seqalign{\span\TL &\quad \span\TT}{
    1 & if $\sgn(-1) = 1$  \cr
    i & if $\sgn(-1) = -1$\,.} \right.
 }
Sometimes we will abbreviate $F^{a_2b_2c_2,a_1b_1c_1}$ to $F^{\cir{2}\cir{1}}$. Graphically, we can keep track of the index structure by representing factors of $\Omega^{a_ib_i}$ with arrows pointing from the second index to the first and using a different color for each type of strand. In this notation, we may represent the free propagator $F^{\cir{2}\cir{1}}$ as
\eqn{propagatordiagram}
{
\begin{matrix} \includegraphics[height=4.1ex]{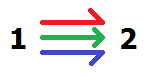} \end{matrix}
,}
and the interaction vertex following from \eqref{Sint} can be represented as
\eqn{vertexdiagram}
{
 \begin{matrix} \includegraphics[height=11ex]{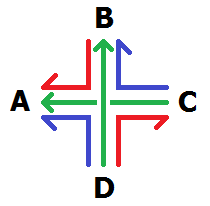} \end{matrix}
.}
Capital letters in \eno{vertexdiagram} and below, like bold numbers in \eno{propagatordiagram}, represent triples of indices: For example, ${\bf A}$ means $a_A b_A c_A$. Note that we may freely flip any arrow at the price of a factor of $\sigma_\Omega$. The fact that the pairwise interchange of all indices in the vertex leaves it invariant up to an even number of arrow flips provides a check that even when $\Omega_{ab}$ is anti-symmetric, the interaction term \eqref{Sint} remains non-vanishing and the interaction vertex retains its tetrahedral symmetry. 

Even when the base field $K$ may no longer be the real numbers, the large $N$ limit with $g^2 N^3$ held fixed is still controlled by the same class of ``melon" diagrams built from the propagator and vertex, irrespectively of whether the theory exhibits $O(N)^3$ or $Sp(N)^3$  symmetry since contracting anti-symmetric versus symmetric matrices $\Omega_{ab}$ at most changes the overall sign of Feynman diagrams and not the scaling.  In particular, the leading correction to the propagator in the melonic limit of $N \to \infty$ with $g^2 N^3$ held fixed is
 \eqn{TwoPtCor}{
  G^{\cir{2}\cir{1}}(t) = F^{\cir{2}\cir{1}}(t) + \delta G^{\cir{2}\cir{1}}(t) \,,
 }
where $\delta G^{\cir{2}\cir{1}}(t)$ is given by the loop diagram 
\eqn{LoopDiagram}
{
 \begin{matrix} \includegraphics[height=20ex]{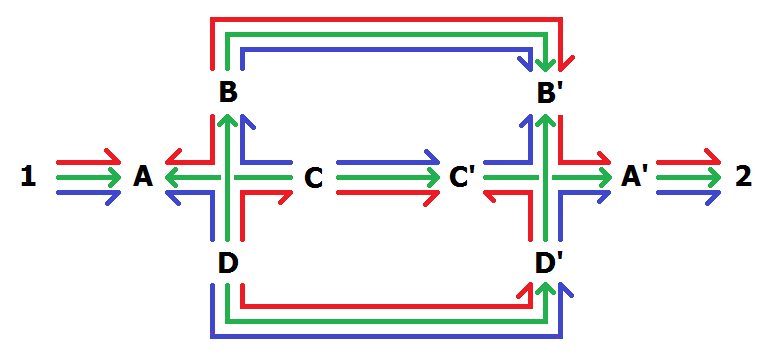} \end{matrix}.
 }
The choice of connecting the index triplets labeled by ${\bf 1}$ and ${\bf A}$ with each other was arbitrary, and likewise the connection between ${\bf A'}$ and ${\bf 2}$. Other choices would lead to diagrams related to the above by flipping an even number of arrows, and all these choices sum together to cancel the factors of $1/4$ coming from \eqref{Sint}. Furthermore, flipping nine arrows, we can turn the loop diagram \eno{LoopDiagram} into
\eqn{LoopDiagram2}
{
 \begin{matrix} \includegraphics[height=20ex]{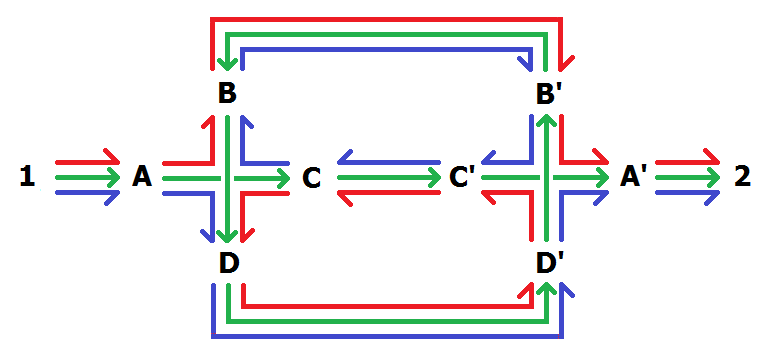} \end{matrix}.
 }
The nine flips cost us a factor of $(\sigma_\Omega)^9=\sigma_\Omega$, but then we have a diagram with all arrows consistently oriented so that the three index loops can each be contracted using \eqref{OmegaInverse} to yield a factor of $N$ each, and the paths going from 1 to 2 can each be contracted to a single factor of $\Omega^{a_ib_i}$. 
Putting everything together, we see that
 \eqn{deltaGagain}{
  \delta G^{a_2b_2c_2,a_1b_1c_1}(t) = \sigma_\Omega g^2 N^3 
    \Omega^{a_2a_1} \Omega^{b_2b_1} \Omega^{c_2c_1} \int_K dt_1 dt_2 \, F(t-t_2) F(t_2-t_1)^3 F(t_1) 
      \,.
 }
 We note from \eno{deltaGagain} that passing from the ${\rm O}(N)$ theory ($\sigma_\Omega=1$) to the ${\rm Sp}(N)$ theory formally corresponds to sending $N \to -N$.  This is reminiscent of the results of \cite{Mkrtchian:1981bb,Cvitanovic:1982bq}; however, those results pertain to the compact ${\rm USp}(N)$ group, whereas our ${\rm Sp}(N)$ is the non-compact group ${\rm Sp}(N,\mathbb{R})$, as mentioned earlier.

We see from \eno{TwoPtCor} and \eno{deltaGagain} that
 \eqn{Gansatz}{
  G^{a_2b_2c_2,a_1b_1c_1}(t) = \Omega^{a_2a_1} \Omega^{b_2b_1} \Omega^{c_2c_1} G(t) \,,
 }
where
 \eqn{GfixedOrder}{
  G(t) = F(t) + \sigma_\Omega g^2 N^3 \int_K dt_1 dt_2 \, F(t-t_2) F(t_2-t_1)^3 F(t_1) + 
    \dots \,,
 }
where the omitted terms are higher order in $g$. 

The standard strategy for finding the conformal behavior in the infrared in SYK-type models is to improve the calculation of $G^{\cir{2}\cir{1}}(t)$ by replacing all but the leftmost propagator in $\delta G$ by $G$ so as to obtain the leading-order Schwinger-Dyson equation,
 \eqn{SDG}{
  G(t) = F(t) + \sigma_\Omega g^2 N^3 \int dt_1 dt_2 \, 
   G(t-t_2) G(t_2-t_1)^3 F(t_1) \,,
 }
and then to argue that \eno{SDG} captures the entire set of diagrams that contributes at leading order in the melonic limit---meaning that omitted diagrams are suppressed by powers of $N$ when we take $N \to \infty$ with $g^2 N^3$ fixed and finite.  Now define an energy scale
 \eqn{muDef}{
  \mu \equiv (g^2 N^3)^{1/(4s-2)} \,.
 }
The claim is that in the infrared, meaning $\mu |t| \gg 1$, the two terms on the right hand side of \eno{SDG} nearly cancel, so that to obtain the leading order expression for $G(t)$ we formally set the left hand side to $0$ and then solve the integral equation to get
 \eqn{SDsoln}{
  G(t) = b {\sgn t \over |t|^{1/2}}
 }
where $b$ is some constant, independent of $t$.  To get hold of this constant, it helps to detour to a more systematic discussion of multiplicative characters, which we undertake in the next section.  We can however extract some information about the phase of $b$ from a general analysis of the reality properties of the Green's function, as follows.  The full Green's function can be expressed as
 \eqn{GEV}{
  G^{a_2b_2c_2,a_1b_1c_1}(t) = \langle \psi^{a_2b_2c_2}(t) \psi^{a_1b_1c_1}(0) \rangle \,,
 }
where $\langle \dots \rangle$ is understood as being defined in terms of a Euclidean path integral based on the real action \eno{KTfull}.  Keeping in mind that $\psi^{abc}$ is itself real and that we are working in Euclidean space, we have
 \eqn{GEVconj}{
  G^{a_2b_2c_2,a_1b_1c_1}(t)^* = \langle \psi^{a_1b_1c_1}(0) \psi^{a_2b_2c_2}(-t) \rangle
   = \sigma_\psi G^{a_2b_2c_2,a_1b_1c_1}(-t) \,.
 }
Recalling that $\Omega^{ab}$ has real entries when $\sigma_\Omega = 1$ and imaginary entries when $\sigma_\Omega = -1$, we conclude that
 \eqn{Gconj}{
  G(t)^* = \sigma_\psi \sigma_\Omega G(-t) = \sgn(-1) G(-t) = G(t) \,.
 }
The second equality of \eno{Gconj} relies on the sign identity deduced from \eno{KTOsym}:
 \eqn{UVsign}{
  \sigma_\psi \sigma_\Omega = \sgn(-1) \,.
 }
We will refer to this identity as the ultraviolet sign constraint because it was forced on us by analysis of the free theory even before we add the relevant interaction term.  By comparing the infrared ansatz \eno{SDsoln} with \eno{Gconj}, we see that $b$ is real.

\section{Multiplicative characters}
\label{MULTIPLICATIVE}

The material in this section is entirely contained in the mathematical literature (see for example \cite{Gelfand:1968}) but has also found applications in $p$-adic physics in the past.\footnote{See \cite{Ruelle:1989jv} for a short discussion of quadratic extensions of the $p$-adics and the $p$-adic sign character in the context of $p$-adic quantum mechanics. 
Multiplicative characters and their associated generalized beta functions have also been studied in the context of non-Archimedean string theory: see \cite{Brekke:1988dg,Arefeva:1988kr,Marshakov:1989jz} for applications to Chan-Paton rules and \cite{Grange:2004xj,Ghoshal:2004ay} for applications to theories with constant Neveu-Schwarz $B$-field.
}  
Some of the explicit results in sections \ref{GammaBeta} and \ref{GammaExplicit} are perhaps harder to find written out in the literature, but they are all elementary applications of the general formalism.

Generally, a multiplicative character of a field is a multiplicative homomorphism $\pi\colon K^\times \to \mathbb{C}^\times$: that is, a map with the property
 \eqn{PiDef}{
  \pi(t_1 t_2) = \pi(t_1) \pi(t_2) \,.
 }
Note that in particular, $\pi(1) = 1$ for all multiplicative characters.\footnote{This is an instance of the general idea that a character is a one-dimensional representation of a group (and so valued in $\text{GL}(1,\mathbb{C}) = \mathbb{C}^\times$): $K^\times$ is regarded as a group under multiplication. 
The collection of multiplicative characters is itself an abelian group, under pointwise multiplication: if $\pi$ and $\pi'$ are multiplicative characters, then so is $(\pi \pi')(x) = \pi(x) \pi'(x)$, and similarly, so is $1/\pi$. (This operation corresponds to tensor product of representations; it is a special property of the one-dimensional representations that the tensor product admits inverses.)} The simplest multiplicative characters are\footnote{Beware a notational conflict with \cite{Gubser:2017vgc}: There $\pi_s(t) = |t|^{s-1}$ in $\mathbb{Q}_p$, whereas here $\pi_s(t) = |t|^s$.}
 \eqn{pisDef}{
  \pi_s(t) = |t|^s \,.
 }
In principle, we can allow any $s \in \mathbb{C}$, but in general we will instead restrict $s \in \mathbb{R}$.  As indicated in \eno{GammaFree}, our central interest is in constructing free propagators from characters of the form 
 \eqn{PiSign}{
  \pi_{s,\sgn}(t) \equiv |t|^s \sgn t \,,
 }
where $\sgn t$ itself is a multiplicative character, taking values in $\{\pm 1\}$ rather than all of $\mathbb{C}^\times$. This is equivalent to asking that the square of the character is the trivial character; such characters are sometimes called ``quadratic.''

In order to give a more precise account of sign characters, we need to understand better the multiplicative structure of the ultrametric fields of interest.  Our main focus is on ultrametric fields $K$ which are finite extensions of $\mathbb{Q}_p$ for some prime $p$.  They are characterized by the prime number $p$, which can be recovered from~$K$ as follows: the residue field $\mathbb{F}_K$ of $K$ has characteristic $p$, meaning that adding the multiplicative identity $1$ to itself $p$ times in~$\mathbb{F}_K$ gives $0$.  ($K$ itself is always of characteristic zero.) To construct $\mathbb{F}_K$, one may first define the ring of integers ${\cal O}_K \equiv \{ t \in K: |t| \leq 1 \}$, and next define $\mathfrak{m}_K \equiv \{ t \in K: |t| < 1 \}$.  Then $\mathfrak{m}_K$ is a maximal ideal in ${\cal O}_K$ (in fact it is the only one), and the quotient $\mathbb{F}_K \equiv {\cal O}_K / \mathfrak{m}_K$ is therefore a field; in fact, it is finite, so that $\mathbb{F}_K = \mathbb{F}_q$, where $q=p^f$ for some positive integer $f$.\footnote{For an unramified extension $K$ of $\mathbb{Q}_p$, $f$ is also the degree of extension of $K$ over $\mathbb{Q}_p$.  In general, $n=ef$ where $e$ is the ramification index, defined as the positive integer such that $|p/\mathfrak{p}^e|_K = 1$ where $\mathfrak{p}$ is a uniformizer for $K$.}

Any element $t \in K$ can be expressed as
 \eqn{tExpress}{
  t = \mathfrak{p}^{v(t)} \epsilon^{w(t)} a(t) \,.
 }
Here $\mathfrak{p}$ is a chosen uniformizer for $K$, i.e., a generator of the maximal ideal $\mathfrak{m}_K$. For example, $\mathfrak{p} = p$ for $\mathbb{Q}_p$, or $\sqrt{p}$ for the totally ramified quadratic extension $\mathbb{Q}_p(\sqrt{p})$. $v(t) \in \mathbb{Z}$ is essentially the valuation; it is defined by the property that
\eqn{valuation}{
|t/\mathfrak{p}^{v(t)}|_K = 1,
}
and is thus proportional to the logarithm of~$|t|$.  $\epsilon$ is a generator of the group $\mathbb{F}_q^\times$ (which is cyclic of order $q-1$), and $w(t) \in \{ 1,2,3,\dots,q-1 \}$ (modulo $q-1$), so that $\epsilon^{w(t)}$ ranges over all elements of $\mathbb{F}_q^\times$.  Finally, $a(t)$ belongs to the multiplicative group
 \eqn{ADef}{
  A = \{ a \in K\colon |a-1|<1 \} \,.
 }
The decomposition \eno{tExpress} is unique once we fix choices for the uniformizer~$\mathfrak{p}$ and the generator~$\epsilon$.  We note that $t \in {\cal O}_K^\times = {\cal O}_K \setminus \mathfrak{m}_K$ (meaning the complement of $\mathfrak{m}_K$ in ${\cal O}_K$) precisely if $v(t) = 0$.  We can think of ${\cal O}_K^\times$ as the analog of the unit circle in $K$ because
 \eqn{UnitCircle}{
  {\cal O}_K^\times = \{ u \in K\colon |u| = 1 \} \,.
 }
Any two uniformizers $\mathfrak{p}$ and $\mathfrak{p}'$ are related by $\mathfrak{p}' = u \mathfrak{p}$ where $u \in {\cal O}_K^\times$.  Thus if we use $\mathfrak{p}'$ instead of $\mathfrak{p}$ in the decomposition \eno{tExpress}, $v(t)$ would remain the same, but $w(t)$ would shift by a fixed multiple of $v(t)$, and $a(t)$ would also change.  Similarly, one can use a different primitive root $\epsilon' = \epsilon^y$, where $y$ is prime to $q-1$, and this would result in multiplying $w(t)$ by $y^{-1}$ in $\mathbb{Z}/(q-1)\mathbb{Z}$.

For future reference, we introduce the notations
 \eqn{ZUnotation}{
  \mathbb{Z}_p \equiv {\cal O}_{\mathbb{Q}_p} \qquad\qquad
  \mathbb{U}_p \equiv {\cal O}_{\mathbb{Q}_p}^\times
 }
for the $p$-adic integers and the $p$-adic units, respectively.

\subsection{Sign characters in finite extensions of $\mathbb{Q}_p$ for odd $p$}

Now let's require $p$ to be an odd prime.  In this case it is fairly straightforward to enumerate the sign characters, because one can show that $\sgn a = 1$ for all $a \in A$.  Thus, the only choices one can make are the values $\sgn\mathfrak{p} = \sigma_{\mathfrak{p}}$ and $\sgn\mathfrak\epsilon = \sigma_\epsilon$, and these choices can be made independently.  In short,
 \eqn{sgnSigma}{
  \rho_{\sigma_{\mathfrak{p}}\sigma_\epsilon}(t) \equiv
    \sigma_{\mathfrak{p}}^{v(t)} \sigma_{\mathfrak\epsilon}^{w(t)} \,.
 }
Changing from one primitive root to another doesn't change the $\sigma_\epsilon^{w(t)}$ factor.  Replacing $\mathfrak{p} \to u \mathfrak{p}$ shifts $w(t) \to w(t) + v(t)$ if the decomposition of $u$ involves an odd power of $\epsilon$; otherwise there is no such shift.  It's clear from \eno{sgnSigma} that there are four independent sign characters, including the trivial one $\rho_{11}$ which assigns $+1$ to all elements of $K$.

We will be particularly interested in sign characters on $K$ which are non-trivial on ${\cal O}_K^\times$, and we will describe such characters as direction-dependent because we can think of the decomposition $t = \mathfrak{p}^{v(t)} u$, where $u \in {\cal O}_K^\times$, as analogous to the polar decomposition $z = r e^{i\theta}$ of a complex number.  We see immediately from \eno{sgnSigma} that direction-dependent sign characters in odd characteristic are precisely the ones where $\sigma_\epsilon = -1$.

\subsection{Sign characters over $\mathbb{Q}_2$}

For $\mathbb{Q}_2$, sign characters are a little more complicated.  The decomposition \eno{tExpress} holds just as before, and the first piece of data one needs to fix a sign character is again $\sgn\mathfrak{p}$.  There are no non-trivial sign characters on $\mathbb{F}_2$ because $\mathbb{F}_2^\times$ is the trivial group consisting of only the identity.  Thus the classification of sign characters over $\mathbb{Q}_2$  comes down to understanding sign characters on the group $A$ defined in \eno{ADef}.  To proceed, we 
choose the uniformizer $\mathfrak{p} = 2$ and express
 \eqn{atExpand}{
  a \equiv a(t) = 1 + 2 a_1 + 4 a_2 + {\cal O}(2^3) \,,
 }
where each $a_i$ is $0$ or $1$, and ${\cal O}(2^3)$ means that the omitted terms are $8$ times a $p$-adic integer.  Then the sign characters on $A$ take the form
 \eqn{rhoA}{
  \rho_{\sigma_{a_1}\sigma_{a_2}}(a) = 
    \sigma_{a_1}^{a_1} \sigma_{a_2}^{a_2} \,,
 }
where each $\sigma_{a_i}$ may be chosen independently to be $\pm 1$.  Thus there are four sign characters on $A$, and consequently eight sign characters
 \eqn{rhoAll}{
  \rho_{\sigma_{\mathfrak{p}}\sigma_{a_1}\sigma_{a_2}}(t) = 
    \sigma_{\mathfrak{p}}^{v(t)} \sigma_{a_1}^{a_1} \sigma_{a_2}^{a_2}
 }
on $\mathbb{Q}_2$, including the trivial one that assigns $+1$ to all $t$.

It is possible to verify that \eno{rhoA} defines a character by direct calculation, but it is helpful to take a more conceptual view.  Let $A^2$ by the group of all elements $t \in \mathbb{Q}_2$ expressible as a square of an element of $A$.  $A^2$ is easily seen to be a subgroup of $A$.  Furthermore, starting from \eno{atExpand} we can easily see that
 \eqn{aSquared}{
  a^2 = 1 + {\cal O}(2^3) \,.
 }
In other words, all elements of $A^2$ have zero second and third digits in their $2$-adic expansion.  It is straightforward to show (using Hensel's Lemma) that this characterization is precise: any element of $A$ with zero second and third digits is in $A^2$.  Therefore the quotient group $A/A^2$ consists of elements precisely of the form \eno{atExpand} (i.e.~each element $A/A^2$ is the set of numbers agreeing with the expansion \eno{atExpand} with fixed $a_1$ and $a_2$).  A convenient characterization of $A/A^2$ is the odd integers modulo $8$ under multiplication, which is isomorphic to ${\mathbb{Z} \over 2\mathbb{Z}} \times {\mathbb{Z} \over 2\mathbb{Z}}$.  Let's write the elements of $A/A^2$ as $\{\pm 1,\pm 3\}$.  The group is generated (multiplicatively modulo $8$) by $\pm 3$, and we can understand $\sigma_{a_1}$ as the sign we choose for $3$ while $\sigma_{a_2}$ is the sign we choose for $-3$.  (This is because $3 =1+2$, while $-3 \equiv 5 = 1 + 2^2$ modulo $8$). The sign of $-1$ is therefore the product $\sigma_{a_1} \sigma_{a_2}$.

Sign characters for extensions of $\mathbb{Q}_2$ are yet more subtle, and we will not have occasion to discuss them explicitly.  They are however well understood: see for example \cite{Hida:2015zz}.

\subsection{An alternative parametrization of sign characters}

An attractive alternative way to parametrize sign characters is to pick an arbitrary element $\tau \in K^\times$ and define
 \eqn{sgnDef}{
  \sgn_\tau t \equiv \left\{ \seqalign{\span\TL &\qquad\span\TT}{
   1 & if $t = a^2 - \tau b^2$ for some $a, b \in K$  \cr
   -1 & otherwise\,.} \right.
 }
Note that $\sgn_\tau = \sgn_{\tau'}$ if $\tau/\tau'$ is a square in $K$.  Thus $\sgn_\tau$ is really parametrized by elements of the finite group $K^\times / (K^\times)^2$, where $(K^\times)^2$ is the set of all elements in $K$ expressible as the square of an element of $K^\times$.  Let's examine cases:
 \begin{itemize}
  \item If $K = \mathbb{R}$, then $\sgn_1$ is the trivial sign character and $\sgn_{-1}$ gives the usual notion of sign in $\mathbb{R}$.  If $K = \mathbb{C}$, then the only possibility is the trivial sign character.
  \item For odd characteristic, $K^\times / (K^\times)^2$ consists of elements $\{1,\mathfrak{p},\epsilon\mathfrak{p},\epsilon\}$.  (More properly, $1$ means $(K^\times)^2$, $\mathfrak{p}$ means $\mathfrak{p} (K^\times)^2$, and so forth.)  The following table provides a translation indicating which $\sgn_\tau$ corresponds to which $\rho_{\sigma_p\sigma_\epsilon}$ for $\mathbb{Q}_p$, where we choose the canonical uniformizer $\mathfrak{p} = p$.  (As noted previously, it doesn't matter which primitive root $\epsilon$ we choose for $\mathbb{F}_p^\times$.)  We also show the value of $\sgn(-1)$, which as seen previously is a key quantity for our analysis.
\newcolumntype{a}{>{\columncolor{Gray}}c}
 \eqn{TauSigmaOdd}{
  \begin{tabular}{|c||c|a|a|c|}
   \hline $\tau$ & $1$ & $p$ & $\epsilon p$ & $\epsilon$ \\ \hline \hline
   $\sigma_p$ & $1$ & $(-1|p)$ & $-(-1|p)$ & $-1$ \\
   $\sigma_\epsilon$ & $1$ & $-1$ & $-1$ & $1$ \\ \hline
   $\sgn_\tau(-1)$ & $1$ & $(-1|p)$ & $(-1|p)$ & $1$ \\ \hline
  \end{tabular}
 }
Here, as above, $(-1|p) = (-1)^{(p-1)/2}$ is the Legendre symbol.  The shaded columns correspond to direction-dependent characters.
  \item $\mathbb{Q}_2^\times / (\mathbb{Q}_2^\times)^2$ consists of elements $\{\pm 1,\pm 2,\pm 3,\pm 6\}$, where factors of $\pm 1$ or $\pm 3$ come from $A/A^2$ 
as discussed above, and the factor of $2$ comes from the uniformizer, which is never a square, and which we choose to be $\mathfrak{p} = 2$. (Some authors prefer to quote the elements of $\mathbb{Q}_2^\times / (\mathbb{Q}_2^\times)^2$ as $\{1,2,3,5,6,7,10,14\}$; these are equivalent presentations.)  The following table provides a translation between $\sgn_\tau$ and $\rho_{\sigma_2\sigma_{a_1}\sigma_{a_2}}$, and it also shows $\sgn(-1)$.
\newcolumntype{b}{>{\columncolor{Gray}}r}
 \eqn{TauSigmaTwo}{
  \begin{tabular}{|c||c|b|b|b|b|c|b|b|}
   \hline $\tau$ & $1$ & $-1$ & $2$ & $-2$ & $3$ & $-3$ & $6$ & $-6$
     \\ \hline \hline
   $\sigma_2$ & $1$ & $1$ & $1$ & $1$ & $-1$ & $-1$ & $-1$ & $-1$ \\ 
   $\sigma_{a_1}$ & $1$ & $-1$ & $-1$ & $1$ & $-1$ & $1$ & $1$ & $-1$ 
     \\
   $\sigma_{a_2}$ & $1$ & $1$ & $-1$ & $-1$ & $1$ & $1$ & $-1$ & $-1$
     \\ \hline
   $\sgn_\tau(-1)$ & $1$ & $-1$ & $1$ & $-1$ & $-1$ & $1$ & $-1$ & $1$ 
     \\ \hline
  \end{tabular}
 }
As before, the shaded columns correspond to direction-dependent characters.
 \end{itemize}
It is worth noting that the elements $\tau \in K^\times/(K^\times)^2$ other than $\tau=1$ also parametrize the possible quadratic extensions $K(\sqrt\tau)$ of $K$; thus each non-trivial sign function on $K$ can be associated with such an extension\footnote{This is the simplest instance of class field theory: quadratic extensions (which are necessarily abelian) correspond to index-two subgroups of~$K^\times$, which are precisely the kernels of sign characters.}---but we should keep in mind that $\sgn_\tau t$ itself takes as its argument $t$ an element of $K^\times$, not the extension $K(\sqrt\tau)$.

\subsection{The generalized $\Gamma$ and ${\rm B}$ functions}
\label{GammaBeta}

Two important quantities for our subsequent analysis are the generalized $\Gamma$ and ${\rm B}$ functions:\footnote{See e.g.~p.~145 of \cite{Gelfand:1968} for a more systematic development; however, beware of typos! See also \cite{Brekke:1993gf}. In general, definitions like \eno{GammaBetaDefs} in terms of integrals work for some range of arguments $\pi$ and $\pi'$; otherwise, we must invoke some process of analytic continuation.}
 \eqn{GammaBetaDefs}{
  \Gamma(\pi) \equiv \int_K {dt \over |t|} \, \chi(t) \pi(t) \qquad\qquad
  {\rm B}(\pi,\pi') \equiv \int_K dt \, {\pi(1-t) \over |1-t|} {\pi'(t) \over |t|} = 
    {\Gamma(\pi) \Gamma(\pi') \over \Gamma(\pi\pi')} \,.
 }
Here $\chi(t)$ is an additive character with $\chi(1) = 1$, while $\pi(t)$ is a multiplicative character; $dt/|t|$ is the multiplicative Haar measure on~$K$.  For $\mathbb{Q}_p$, we have $\chi(t) = e^{2\pi i \{ t \}}$, where $\{ t \}$ denotes the fractional part of $t$.  For extensions of $\mathbb{Q}_p$, the additive character still has the intuitive interpretation of a plane wave; see for example \cite{Gubser:2016guj} for an explicit discussion of the case of an unramified extension.\footnote{
For extensions of the $p$-adic field, one can define
 \eqn{zetaGammaDef}{
  \zeta_q(s) \equiv {1 \over 1 - q^{-s}} \qquad\qquad
  \Gamma_q(s) \equiv {\zeta_q(s) \over \zeta_q(1-s)} \qquad\qquad
  {\rm B}_q(s_1,s_2) \equiv {\Gamma_q(s_1) \Gamma_q(s_2) \over \Gamma_q(s_1+s_2)}
 }
for any $q = p^n$ where $p$ is prime and $n$ is a positive integer. In the unramified extension $\mathbb{Q}_q$, we then have
 \eqn{UnramComparison}{
  \Gamma(\pi_s) = \Gamma_q(s) \qquad\qquad
  {\rm B}(\pi_{s_1},\pi_{s_2}) = {\rm B}_q(s_1,s_2) \,.
 }
Note the contrast with \cite{Gubser:2016guj,Gubser:2017vgc}, where we eschewed defining $\zeta_q$ and worked only with $\zeta_p$. 
}

There are two obvious ways to combine multiplicative characters.  First, as mentioned above, we may multiply pointwise:
 \eqn{PiMultiply}{
  (\pi\pi')(t) \equiv \pi(t) \pi'(t) \,.
 }
Second, we may use convolutions, defined as
 \eqn{ConvolveDef}{
  (f*g)(t) = \int dt_1 \, f(t-t_1) g(t_1) \,.
 }
A useful identity is
 \eqn{ConvolveID}{
  (\pi * \pi')(t) = {\rm B}(\pi\pi_1,\pi'\pi_1) (\pi\pi'\pi_1)(t) \,.
 }
Here, $\pi_1$ refers to the character $\pi_s = |\cdot|^s$ defined in \eno{pisDef}, with $s=1$. This is closely related to another useful identity:
 \eqn{FourierID}{
  \int dt \, \chi(\omega t) \pi(t) = \Gamma(\pi\pi_1) (\pi^{-1}\pi_{-1})(\omega) \,,
 }
where $\pi^{-1}$ indicates the pointwise multiplicative inverse: $\pi^{-1}(t) \equiv 1/\pi(t)$.  One more useful identity can be obtained by noting that
 \eqn{InverseFourier}{
  \int d\omega \, \chi(-\omega t) \int dt' \, \chi(\omega t') f(t') = f(t) \,.
 }
Applying this to \eno{FourierID} gives 
 \eqn{GammaInversion}{
  \Gamma(\pi\pi_1) \Gamma(\pi^{-1}) = \pi(-1) \,, }
and we note that $\pi(-1) = \pm 1$ since $(-1)^2 = 1$; in particular, $\pi_{s,\sgn}(-1) = \sgn(-1)$.
The reader may wonder how to reconcile our conventions for gamma functions with the intuition that $\Gamma$ is typically a function of a complex variable. If we restrict to characters of the form~$\pi_s$, where $s\in \mathbb{C}$, the usual intuition is recovered. In particular, \eno{ConvolveID} says that this family of characters (which is additive pointwise, in the sense that $\pi_s\pi_t = \pi_{s+t}$) is also additive under convolutions, up to a scaling and shift that were introduced by our choice of conventions: It specializes to
\eqn{AdditiveConvolution}{
\pi_s * \pi_t = {\rm B}(\pi_{s+1},\pi_{t+1}) \pi_{s+t+1}.
}
We can now demonstrate that the free action \eno{Sfree} is bilocal in position space. Using \eno{FourierID} we may Fourier transform \eno{Sfree} to obtain
 \eqn{SfreePos}{
  S_{\rm free} =  {1 \over \Gamma(\pi_{-s,\sgn})} \int_K dt_1 dt_2 \, 
   {1 \over 2} \psi^{a_1 b_1 c_1}(t_1) \Omega_{a_1 a_2} \Omega_{b_1 b_2} \Omega_{c_1 c_2}
    {\sgn(t_1-t_2) \over |t_1 - t_2|^{1+s}}
     \psi^{a_2 b_2 c_2}(t_2)  \,,
 }
 where $\pi_{s,\sgn}$ is defined in \eno{PiSign}.
 The non-local operator in \eno{SfreePos} appears in place of a local time derivative in standard Archimedean field theories, and is closely related to the generalized Vladimirov derivative,
 \eqn{Vlad}{
 D^{s,\sgn} \phi (t) \equiv {1 \over \Gamma(\pi_{-s,\sgn})} \int dt^\prime\, {\sgn(t-t^\prime) \over |t-t^\prime |^{1+s}} \left(\phi(t^\prime) - \phi(t)\right).
 }
 This relation is explained in detail in the appendix of \cite{Gubser:2016htz} for the usual Vladimirov derivative with a trivial sign character, $D^s$,  but carries over just as well in the case of the generalized Vladimirov derivative $D^{s,\sgn}$.

\subsection{Generalized $\Gamma$ functions for $\mathbb{R}$ and $\mathbb{Q}_p$}
\label{GammaExplicit}

For $K=\mathbb{R}$ and for $K=\mathbb{Q}_p$, we want to evaluate the generalized gamma function $\Gamma$ on characters of the form
 \eqn{characterdef}{
  \pi_{s,\tau}(t) \equiv |t|^s \sgn_\tau t \,,
 }
where $s \in \mathbb{R}$ while $\tau$ and $t$ are in $K^\times$.  It is useful first to define local zeta functions
 \eqn{zetaDefs}{
  \zeta_\infty(s) \equiv \pi^{-s/2} \Gamma_{\rm Euler}(s/2) \qquad\qquad
   \zeta_p(s) = {1 \over 1-p^{-s}} \,,
 }
where $\Gamma_{\rm Euler}(z)$ is the usual Euler gamma function.  Note that over $\mathbb{R}$, the generalized gamma function is different from $\Gamma_{\rm Euler}$, though related:
 \eqn{GammaReal}{
  \Gamma(\pi_{s,1}) &= \int_{-\infty}^\infty dt \, |t|^{s-1} e^{2\pi i t} = 
    {\zeta_\infty(s) \over \zeta_\infty(1-s)} = 
     {2 \cos {\pi s \over 2} \over (2\pi)^s} \Gamma_{\rm Euler}(s)  = {\zeta(1-s) \over \zeta(s)} \cr
  \Gamma(\pi_{s,-1}) &= \int_{-\infty}^\infty dt \, |t|^{s-2} t \, e^{2\pi i t} = {-1 \over 2 i 4^{s-1} }\frac{\zeta_\infty(1-s) \zeta_\infty(2 s)}{\zeta_\infty(s) \zeta_\infty(2-2 s)}
    = 
     {2i \sin {\pi s \over 2} \over (2\pi)^s} \Gamma_{\rm Euler}(s) \cr 
     &= {-1 \over 2i 4^{s-1}} {L(1-s, \chi) \over L(s,\chi)}\,,
 }
where $\zeta(s)$ is the usual Riemann zeta function,
 \eqn{RiemannZeta}{
 \zeta(s) = \sum_{n=1}^\infty {1 \over n^s} = \prod_{p}{1 \over 1-p^{-s}} = \prod_p \zeta_p(s)\,,
 }
and $L(s,\chi)$ is a Dirichlet $L$-function,
  \eqn{Lfn}{
 L(s,\chi) = \sum_{n=1}^\infty {\chi(n) \over n^s} = \prod_{p}{1 \over 1-\chi(p)p^{-s}} \,.
 }
The particular Dirichlet character $\chi$ that appears in \eno{GammaReal} and \eno{Lfn} has modulus four; it is defined on primes through $\chi(2) = 0$ and $\chi(p) = (-1|p)$ for odd primes.  Its values on all positive integers are then fixed by the multiplicative property, $\chi(mn) = \chi(m) \chi(n)$ for positive $m$ and $n$ \cite{Ruelle:1989dg}. One can think of it as defined by pulling back the unique non-trivial character of $(\mathbb{Z}/4\mathbb{Z})^\times$ to~$\mathbb{Z}$ along the quotient map, after extending by zero. We will not encounter Dirichlet characters elsewhere in the paper.  We will instead use $\chi$ to denote an additive character on a field $K$.

Over $\mathbb{C}$, as explained above, only the trivial sign character is available, and we have
\eqn{ComplexGamma}
{\Gamma_{\mathbb{C}}(\pi_s)=\int_{\mathbb{C}} dz\, e^{2\pi i(z+\bar{z})}|z|^{s-1}=(2\pi)^{-2s}[\Gamma_{\text{Euler}}(s)]^2\sin(\pi s).
 }
Now let's consider $\mathbb{Q}_p$ for odd primes $p$.  For the trivial sign function, using equation (22) of \cite{Gubser:2016guj}, we see that
\eqn{trivsigngamma}
{
\Gamma(\pi_{s,1}) =\sum_{\ell=-\infty}^\infty (p^{-s})^\ell \int_{\mathbb{U}_p}dt\, \chi(p^\ell t)=\sum_{\ell=0}^\infty(p^{-s})^\ell-\frac{1}{p}\sum_{\ell=-1}^\infty(p^{-s})^\ell=\frac{\zeta_p(s)}{\zeta_p(1-s)} \,,
}
which is a standard result; note the similarity to the first line of \eno{GammaReal}.  For non-trivial sign functions, calculations similar to \eno{trivsigngamma} can be carried out with the help of the table in \eno{TauSigmaOdd}.  For example: When $\tau=\epsilon$, we see from the table that $\sgn_\tau t$ depends only on the parity of $v(t)$, the above sum becomes alternating, and the gamma function evaluates to 
\eqn{gammaecomp}
{
\Gamma(\pi_{s,\epsilon}) =\sum_{\ell=-\infty}^\infty (-p^{-s})^\ell \int_{\mathbb{U}_p}dt\, \chi(p^\ell t)=\sum_{\ell=0}^\infty(-p^{-s})^\ell-\frac{1}{p}\sum_{\ell=-1}^\infty(-p^{-s})^\ell = \frac{\zeta_p(1-s)\zeta_p(2s)}{\zeta_p(s)\zeta_p(2-2s)} \,.
}
 Note the similarity of \eno{gammaecomp} to the second line of \eno{GammaReal}.
For $\tau=p$ and $\tau=\epsilon p$, we have
\eqn{vanishint}
{\int_{\mathbb{U}_p}dt\,\chi(p^\ell t)\sgn_\tau t = 0 \qquad
  \hbox{for integers $\ell \neq -1$} \,,}
because in these cases the range of the additive character is symmetrically distributed around the unit circle. The case $\ell = -1$ can be dealt with by invoking the Gauss sum
\eqn{Gausssum}{
\sum_{k=1}^{p-1}e^{\frac{2\pi i k}{p}}(k|p)=\begin{cases} \sqrt{p}, \hspace{7mm} p\equiv 1 \mod 4,
\\ i\sqrt{p}, \hspace{5mm} p\equiv 3 \mod 4, \end{cases}}
 and so the gamma function in these cases is given by
\eqn{gammaepcomp}
{
\Gamma(\pi_{s,\tau}) &=\sum_{\ell=-\infty}^\infty (p^{-s})^\ell (\sgn_\tau p)^\ell \int_{\mathbb{U}_p}dt\, \chi(p^\ell t)\sgn_\tau t  \cr
 &= 
\begin{cases}
p^{s-\frac{1}{2}}, \hspace{10mm} \tau=p, \hspace{4mm} p\equiv 1 \mod 4
\\
-i p^{s-\frac{1}{2}},  \hspace{5.5mm} \tau=p, \hspace{4mm} p\equiv 3 \mod 4
\\
-p^{s-\frac{1}{2}},  \hspace{7mm} \tau=\epsilon p, \hspace{2mm} p\equiv 1 \mod 4
\\
i p^{s-\frac{1}{2}},  \hspace{9mm} \tau=\epsilon p, \hspace{2mm} p\equiv 3 \mod 4 \,.
\end{cases}
}
Turning now to the gamma functions in the field $\mathbb{Q}_p$ when $p=2$, we start by noticing that the gamma functions for the trivial sign and for $\tau=-3$ are respectively equal to \eqref{trivsigngamma} and \eqref{gammaecomp} with $p=2$. For $\tau=-1$ and $\tau=3$, the sign function depends only on the 2-adic digit $a_1$; decomposing the integral over $\mathbb{Q}_2$ into a sum of integrals over $2^\ell\mathbb{U}_2$, only the $\ell=-2$ term is non-vanishing, and within the domain $\mathbb{U}_2/2^2$ the additive character $\chi(t)$ also only depends on $a_1$ so that the integral can be split into two parts with constant integrands. The volume of the domain $\mathbb{U}_2/2^2$ is twice as large as that of $\mathbb{Z}_2$, which is normalized to unity. We therefore have
\eqn{p2first}{
\Gamma(\pi_{s,\tau})=\int_{\mathbb{U}_2/2^2}dt\, |t|^{s-1} \chi(t)\sgn_\tau t=4^{s-1}\bigg(e^{2\pi i\,\frac{1}{4}}-e^{2\pi i\,\frac{3}{4}}\bigg)=\frac{i}{2}4^s.
}
When $\tau$ is equal to $\pm 2$ or $\pm 6$, $\sgn_\tau t$ depends on both $a_1$ and $a_2$, and in the decomposition of $\mathbb{Q}_p$ into a sum over $2^\ell\mathbb{U}_2$, it is the $\ell=-3$ term that is non-vanishing. For $\tau$ equal to $-2$ or $6$, $\sgn_\tau t$ doesn't depend on $a_1$, and we have
\eqn{p2second}{
\Gamma(\pi_{s,\tau})=\int_{\mathbb{U}_2/2^3}dt \, |t|^{s-1} \chi(t)\sgn_\tau t=\sigma_2\, 8^{s-1}\bigg(e^{2\pi i\,\frac{1}{8}}+e^{2\pi i\,\frac{3}{8}}-e^{2\pi i\,\frac{5}{8}}-e^{2\pi i\,\frac{7}{8}}\bigg)=\frac{i\sigma_2}{\sqrt{8}}8^s.}
For $\tau$ equal to $2$ or $-6$, $\sgn_\tau t$ depends on $a_1$ as well as $a_2$, and we have
\eqn{p2third}{
\Gamma(\pi_{s,\tau})=\int_{\mathbb{U}_2/2^3}dt\,|t|^{s-1}\chi(t)\sgn_\tau t=\sigma_2\, 8^{s-1}\bigg(e^{2\pi i\,\frac{1}{8}}-e^{2\pi i\,\frac{3}{8}}-e^{2\pi i\,\frac{5}{8}}+e^{2\pi i\,\frac{7}{8}}\bigg)=\frac{\sigma_2}{\sqrt{8}}8^s.
}
The dependence on $\sigma_2$ in equations \eqref{p2second} and \eqref{p2third}, in contradistinction to equation \eqref{p2first}, is due to the fact that $v(t)$ is odd for $t\in 2^{-3}\mathbb{U}_2$, unlike the case when $t\in 2^{-2}\mathbb{U}_2$.

An important point to keep in mind is that precisely for direction-dependent sign characters, we find a pure power dependence of the generalized gamma function on $s$: that is, $\Gamma(\pi_{s,\tau}) = K e^{\kappa s}$ for some constants $K$ and $\kappa$ which depend on the base field and $\tau$, but not on $s$.  This makes sense because the integral defining the generalized gamma function can be restricted to a scaled copy of $\mathbb{U}_p$ in these cases.  By way of contrast, for direction-independent characters the integral can be restricted only to a scaled copy of $\mathbb{Z}_p$, and the infinite geometric sums that give more complicated $s$ dependence arise from splitting this copy of $\mathbb{Z}_p$ into a semi-infinite collection of copies of $\mathbb{U}_p$.

\section{Solving the Schwinger-Dyson equation}
\label{SOLUTION}

With a fuller account of multiplicative characters now in hand, let's revisit the Schwinger-Dyson equation \eno{SDG}.  It can be expressed using convolutions as
 \eqn{SDGagain}{
  G = F + \sigma_\Omega g^2 N^3 (G * G^3 * F) \,,
 }
where $G$ and $F$ refer to the position space Green's functions $G(t)$ and $F(t)$.  In section~\ref{INFRARED}, we will solve \eno{SDGagain} in the infrared limit and show that for each choice of field $K$ and each choice of character, a combination of constraints from the ultraviolet and infrared behavior forces us to choose particular signs $\sigma_\Omega$ and $\sigma_\psi$: that is, we have only one option in each case regarding the symmetry group and the statistics of the field $\psi$.

\subsection{Infrared limit}
\label{INFRARED}

The strategy for finding the infrared behavior is to regard $g^2 N^3$ as large and recognize that, to leading order in inverse powers of $g^2 N^3$, we should set $G=0$ on the left hand side of \eno{SDGagain} and then solve the resulting equation for $G$.  More precisely, we have 
 \eqn{Fform}{
  F(t) = \,  \frac{1}{{\sqrt{\sgn(-1)}}}\,\Gamma(\pi_{1-s,\sgn}) \pi_{s-1,\sgn}(t)
 }
by Fourier transforming \eno{FwForm} with the help of \eno{FourierID}, and we propose
 \eqn{Gform}{
  G(t) = b \pi_{-{1 \over 2},\sgn}(t)
 }
in the infrared. The notation $\pi_{s,\sgn}$ for the multiplicative character, as introduced in \eno{PiSign}, is used to stress that we are not committing to any specific $\tau$ in \eno{characterdef}. Of course, \eno{Gform} is just a rewriting of \eno{SDsoln}.

Starting from \eno{Fform} and \eno{Gform} we note first that $G(t)^3 = b^3 \pi_{-{3 \over 2},\sgn}(t)$, and next that \eno{SDGagain} with the left hand side set to zero simplifies to
 \eqn{Pid}{
  \pi_{-{3 \over 2},\sgn} * \pi_{-{1 \over 2},\sgn} * \pi_{s-1,\sgn} = 
    -{1 \over b^4 \sigma_\Omega g^2 N^3} \pi_{s-1,\sgn} \,,
 }
which we recognize as an eigenvalue equation. In fact, it is a totally degenerate eigenvalue equation: by identities such as~\eno{ConvolveID}, the operator on the left-hand side is proportional to the identity, and the eigenvalue equation reduces to a single identity between numbers: all dependence on the spectral parameter~$s$ simply cancels out.
To say this another way, we can pass to momentum space, because a Fourier transform converts the convolutions in~\eno{Pid} into products.  Using
 \eqn{SpecialSDcases}{
  \int dt \, \chi(\omega t) \pi_{-{3 \over 2},\sgn}(t) &= 
    \Gamma(\pi_{-{1 \over 2},\sgn}) \pi_{{1 \over 2},\sgn}(\omega)  \cr
  \int dt \, \chi(\omega t) \pi_{-{1 \over 2},\sgn}(t) &=
    \Gamma(\pi_{{1 \over 2},\sgn}) \pi_{-{1 \over 2},\sgn}(\omega) \,,
 }
both of which are special cases of \eno{FourierID}, we see immediately that in momentum space \eno{Pid} simplifies to
 \eqn{PidOmega}{
  \Gamma(\pi_{-{1 \over 2},\sgn}) \Gamma(\pi_{{1 \over 2},\sgn}) \pi_{-s,\sgn} = 
    {1 \over b^4 \sigma_\Omega g^2 N^3} \pi_{-s,\sgn} \,.
 }
Thus we arrive at
 \eqn{PidSolve}{
  {1 \over b^4 g^2 N^3} = -\sigma_\Omega \Gamma(\pi_{-{1 \over 2},\sgn})
    \Gamma(\pi_{{1 \over 2},\sgn}) \,.
 }
The left hand side of \eno{PidSolve} is positive because $b$ is real (recall the discussion at the end of section~\ref{DIAGRAMS}).  Once we fix $\sgn$, we must then choose $\sigma_\Omega$ in order to make the right hand side of \eno{PidSolve} positive.  If for some reason this cannot be done, then the proposed infrared scaling \eno{Gform} is impossible, and something else must happen to the theory \eno{KTfull} in the infrared.  Having fixed $\sigma_\Omega$ as we just described, we next use the ultraviolet constraint \eno{UVsign} to arrive at a definite choice of $\sigma_\psi$.  In short, we find that {\it for each field $K$ (Archimedean or ultrametric) and each sign character (trivial or non-trivial), there can only be one class of theories of the type \eno{KTfull}, specified by a choice of $\sigma_\Omega$ and $\sigma_\psi$ and parametrized by $s \in (1/2,1]$, which exhibits scaling behavior of the type \eno{Gform} in the infrared.}

Four points are worth noting:
 \begin{itemize}
  \item So far in this section we have not actually used any detailed knowledge of multiplicative characters.  Where this knowledge comes into play is in evaluating $\Gamma(\pi_{-{1 \over 2},\sgn}) \Gamma(\pi_{{1 \over 2},\sgn})$.
  \item For each fixed value of $s \in (1/2,1]$, we have a renormalization group flow from a free theory in the ultraviolet (deformed by the interaction term, which is relevant) to the scaling \eno{Gform} in the infrared.  Allowing $s$ to vary, we see that the class of theories we get for definite $K$ and $\sgn$ is a line of free ultraviolet theories, all of which exhibit the same scaling behavior in the infrared when deformed by the relevant interaction \eno{Sint}.
  \item The result \eno{PidSolve} makes clear that even the normalization $b$ of the two-point function does not depend on $s$ (though it clearly does depend on $K$ and $\sgn$).  This is a curiously strong form of universality for which we do not have a simple intuitive explanation.
  \item The general analysis at the end of section~\ref{DIAGRAMS} fixes the phase of $b$ up to a sign.  Nothing in the infrared analysis we have presented in this section resolves the sign ambiguity in $b$; see however section~\ref{EXACT} for the case of direction-dependent characters.
 \end{itemize}

Passing to $K = \mathbb{R}$ or $\mathbb{Q}_p$, we can use the results of section~\ref{MULTIPLICATIVE} to give a fully explicit account of which theories exist in which places.  The results are shown in table~\ref{ExplicitResults}.  Some further remarks are in order:
\definecolor{Gray}{gray}{0.9}
\def\gr{\rowcolor{Gray}}
\renewcommand{\arraystretch}{1.3}
 \begin{table}\begin{center}\begin{tabular}{|c||c|r|c|r|r|l|}
  \hline
  $K$ & condition & $\tau$ & 
    $\Gamma(\pi_{-{1 \over 2},\sgn}) \Gamma(\pi_{{1 \over 2},\sgn})$ &
    $\sigma_\Omega$ & $\sigma_\psi$ & explanation \\[3pt]
    \hline\hline
  $\mathbb{R}$ & & $1$ & $-4\pi$ & $1$ & $1$ & ${\rm O}(N)$ bosonic \\
  $\mathbb{R}$ & & $-1$ & $-4\pi$ & $1$ & $-1$ & ${\rm O}(N)$ fermionic  \\ \hline
  $\mathbb{C}$ & & $1$ & $-4\pi^2$ & $1$ & $1$ & ${\rm O}(N)$ bosonic  \\ \hline
  $\mathbb{Q}_p$ & $p$ odd & $1$ & $-(p+\sqrt{p}+1)/p^{3/2}$ & $1$ & $1$ & ${\rm O}(N)$ bosonic  \\
  $\mathbb{Q}_p$ & $p$ odd & $\epsilon$ & $(p-\sqrt{p}+1)/p^{3/2}$ & $-1$ & $-1$ & ${\rm Sp}(N)$ fermionic  \\ 
  \gr $\mathbb{Q}_p$ & $p \equiv 1 \mod 4$ & $p$ & $1/p$ & $-1$ & $-1$ & ${\rm Sp}(N)$ fermionic  \\ 
  \gr $\mathbb{Q}_p$ & $p \equiv 1 \mod 4$ & $\epsilon p$ & $1/p$ & $-1$ & $-1$ & ${\rm Sp}(N)$ fermionic  \\
  \gr $\mathbb{Q}_p$ & $p \equiv 3 \mod 4$ & $p$ & $-1/p$ & $1$ & $-1$ & ${\rm O}(N)$ fermionic  \\
  \gr $\mathbb{Q}_p$ & $p \equiv 3 \mod 4$ & $\epsilon p$ & $-1/p$ & $1$ & $-1$ & ${\rm O}(N)$ fermionic  \\ \hline
  $\mathbb{Q}_2$ & & $1$ & $-(2+3\sqrt{2})/4$ & $1$ & $1$ & ${\rm O}(N)$ bosonic  \\
  \gr $\mathbb{Q}_2$ & & $-1$ & $-1/4$ & $1$ & $-1$ & ${\rm O}(N)$ fermionic  \\
  \gr $\mathbb{Q}_2$ & & $2$ & $1/8$ & $-1$ & $-1$ & ${\rm Sp}(N)$ fermionic  \\
  \gr $\mathbb{Q}_2$ & & $-2$ & $-1/8$ & $1$ & $-1$ & ${\rm O}(N)$ fermionic  \\
  \gr $\mathbb{Q}_2$ & & $3$ & $-1/4$ & $1$ & $-1$ & ${\rm O}(N)$ fermionic  \\
  $\mathbb{Q}_2$ & & $-3$ & $(3-\sqrt{2})/\sqrt{8}$ & $-1$ & $-1$ & ${\rm Sp}(N)$ fermionic  \\
  \gr $\mathbb{Q}_2$ & & $6$ & $-1/8$ & $1$ & $-1$ & ${\rm O}(N)$ fermionic  \\
  \gr $\mathbb{Q}_2$ & & $-6$ & $1/8$ & $-1$ & $-1$ & ${\rm Sp}(N)$ fermionic  \\ \hline
 \end{tabular}\end{center}\caption{Explicit characterizations of melonic theories exhibiting renormalization group flows from free theories in the ultraviolet to a strongly interacting fixed point in the infrared.  An exact solution of the renormalization group flow is available for the shaded rows, as explained below.}\label{ExplicitResults}\end{table}

 \begin{itemize}
  \item Fermionic theories are required whenever the sign character is non-trivial; otherwise we need bosonic theories.
  \item The fermionic Klebanov-Tarnopolsky model corresponds to the second line of the table.
  \item The bosonic Klebanov-Tarnopolsky models in one and two dimensions correspond to the first and third lines of the table.  (More properly, these rows correspond to modified bosonic Klebanov-Tarnopolsky models with bilocal kinetic terms; c.f.~footnote \ref{LowerCriticalDim}.)  Higher-dimensional bosonic models are not part of our story because we insist that the theory be defined over a field rather than a vector space.  A systematic extension to vector spaces over fields both Archimedean and ultrametric seems like a large but possibly important undertaking.
 \end{itemize}

\subsection{An exact solution of the Schwinger-Dyson equation}
\label{EXACT}

In general, the Schwinger-Dyson equation is difficult to solve because it involves non-linear combinations of the dressed propagator $G$ (which is what we have to solve for) and must be expressed as a combination of ordinary products and convolutions---so it is local neither in position space nor in momentum space.  But for direction-dependent ultrametric characters, there is a way to beat this: It turns out that for these characters, the Schwinger-Dyson equation is local both in position space and in momentum space, so we can solve it (in either position space or momentum space) just by solving a quartic equation at each energy scale.

The key to this striking result is a Fourier transform identity on $\mathbb{Q}_p$, most easily stated for the case of odd $p$, where the direction-dependent sign characters are $\sgn_\tau$ for $\tau=p$ and $\epsilon p$:
 \eqn{StrikingFourier}{
  \int_{\mathbb{Q}_p} dt \, \chi(\omega t) \delta_{v(t)} \sgn t = 
   M_{-1} \delta_{v(\omega)+1} \sgn\omega \,,
 }
where we are restricting to direction-dependent sign characters (i.e.~$\sgn = \sgn_p$ or $\sgn_{\epsilon p}$), and we remind the reader that $v(t)$ is the valuation of $t$ (see the discussion around \eno{valuation}).  We define
 \eqn{deltaDef}{
  \delta_v \equiv \left\{ \seqalign{\span\TL &\quad\span\TT}{
    1 & if $v=0$  \cr  0 & otherwise} \right.
 }
to be the characteristic function of~$\mathbb{U}_p$, and
 \eqn{Mexpress}{
  M_{-1} \equiv \frac{\Gamma(\pi_{1,\sgn})}{p} = 
   \pm \sqrt{\sgn(-1) \over p} \,.
 }
The sign in the last expression can be read off from \eno{gammaepcomp}.  Indeed, \eno{StrikingFourier} is just a rephrasing of \eno{vanishint}-\eno{gammaepcomp}.

\begin{figure}[h]\centerline{\includegraphics[width=6in]{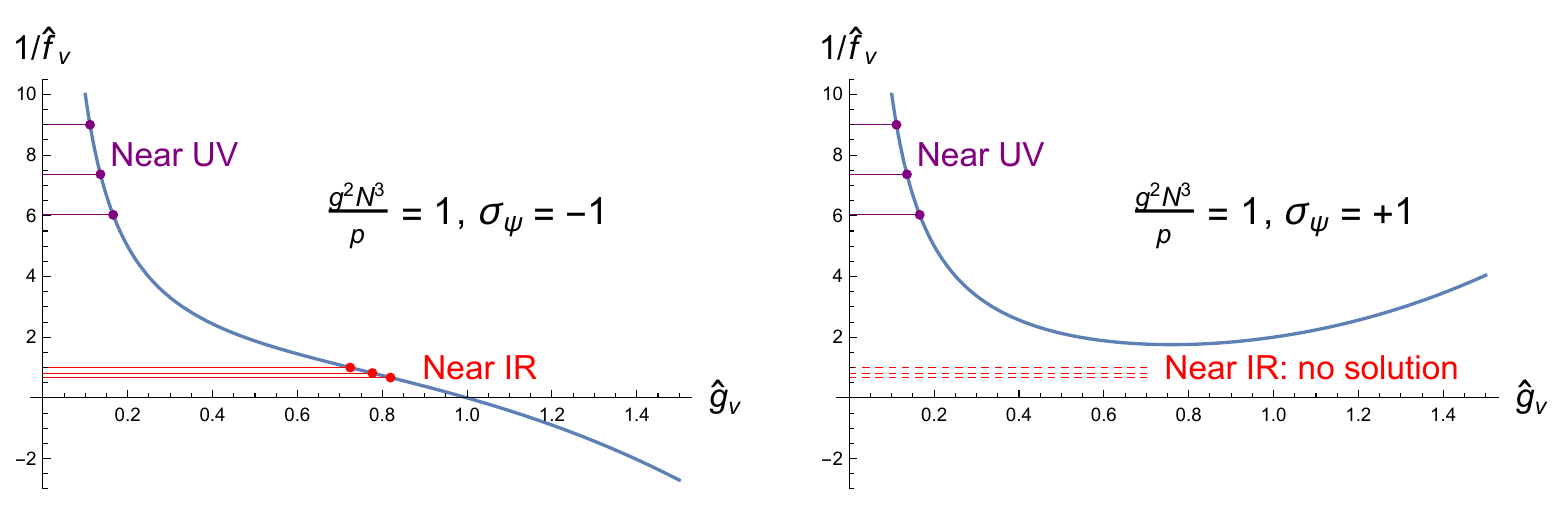}}
  \caption{Solving the quartic equation \eno{QPagain}.  We use the value ${g^2 N^3 \over p}=1$ as an example in order to be able to draw a definite curve of $1/\hat{f}_v$ versus $\hat{g}_v$.  Other positive values of ${g^2 N^3 \over p}$ give qualitatively similar results.  Suppose we pick a value of $\hat{f}_v$: small if we're in the UV, and large in the IR.  If $\sigma_\psi = -1$, as in the left-hand plot, then there is always a unique positive solution $\hat{g}_v$ corresponding to the chosen value of $\hat{f}_v$.  But if $\sigma_\psi = +1$, as in the right-hand plot, then sufficiently far into the IR we have no such solution, and it appears therefore that the interacting theory is ill-defined.}\label{QuarticSoln}
 \end{figure}

Now let's make a more general ansatz than \eno{Gform} for $G(t)$, allowing completely general dependence on the magnitude of~$t$ so that we may write
 \eqn{FandG}{
  F(t) = f_{v(t)} \sgn t \qquad\qquad 
  G(t) = g_{v(t)} \sgn t \,.
 }
Straightforward application of \eno{StrikingFourier} leads to
 \eqn{FandGfourier}{
  F(\omega) &= M_{-1} p^{v(\omega)+1} f_{-v(\omega)-1} \sgn\omega  \cr
  G(\omega) &= M_{-1} p^{v(\omega)+1} g_{-v(\omega)-1} \sgn\omega \,,
 }
and if we define $H(t) \equiv G(t)^3 = g_{v(t)}^3 \sgn t$, then we have also
 \eqn{Hfourier}{
  H(\omega) &= M_{-1} p^{v(\omega)+1} g_{-v(\omega)-1}^3 \sgn\omega \,.
 }
In momentum space, the Schwinger-Dyson equation \eno{SDGagain} reads   
 \eqn{SDGfourier}{
  G(\omega) = F(\omega) + \sigma_\Omega g^2 N^3 G(\omega) H(\omega) F(\omega) \,.
 }
Plugging \eno{FandGfourier} and \eno{Hfourier} into \eno{SDGfourier}, and then simplifying using \eno{Mexpress} and the ultraviolet sign constraint \eno{UVsign}, one arrives at
 \eqn{QuarticPolynomial}{
  g_v = f_v + {\sigma_\psi g^2 N^3 \over p} p^{-2v} g_v^4 f_v \,.
 }
Referring to \eno{Fform}, we see that we can express
 \eqn{fForm}{
  f_v = \theta p^{v/2} \hat{f}_v \qquad\hbox{where}\qquad 
    \hat{f}_v = p^{\left( {1 \over 2} - s \right) (v+1)} \,,
 }
and $\theta$ is equal to plus or minus one.
By design, $\hat{f}_v$ is positive.  Note also that $\hat{f}_v$ becomes small in the ultraviolet and large in the infrared.  Let us likewise express
 \eqn{gForm}{
  g_v = \theta p^{v/2} \hat{g}_v \,.
 }
The reality property $G(t)^* = G(t)$ demonstrated in \eno{Gconj} forces $\hat{g}_v \in \mathbb{R}$.  Using \eno{fForm}-\eno{gForm}, \eno{QuarticPolynomial} can be re-expressed as
 \eqn{QPagain}{
  {1 \over \hat{f}_v} = {1 \over \hat{g}_v} + {\sigma_\psi g^2 N^3 \over p} \hat{g}_v^3 \,,
 }
and this form has the advantage that there is no explicit $v$ dependence in the coefficients of the relation between $\hat{f}_v$ and $\hat{g}_v$.  It is easy to show (see figure~\ref{QuarticSoln}) that there is a unique positive real solution $\hat{g}_v$ to \eno{QPagain} for each $\hat{f}_v$, provided $\sigma_\psi = -1$, and in the infrared, where $\hat{f}_v$ becomes large, $\hat{g}_v$ converges to a finite answer:
 \eqn{IRgv}{
  \hat{g}_v \to \left( {p \over g^2 N^3} \right)^{1/4} \qquad\hbox{as}\qquad
     v \to -\infty \,.
 }
Note that the result \eno{IRgv} is consistent with the general infrared result \eno{PidSolve} given the simple values of $\Gamma(\pi_{\pm {1 \over 2},\sgn})$ for direction-dependent characters.  We must choose $\hat{g}_v$ positive when solving \eno{QPagain}, because for fixed $v$, $\hat{g}_v$ must interpolate smoothly from $\hat{f}_v$ to $\left( {p \over g^2 N^3} \right)^{1/4}$ as we vary the continuous parameter $g^2 N^3$ from small values to large values.  This amounts to a determination of the sign of $b$ in the infrared.

If we pick $\sigma_\psi = +1$ in \eno{QPagain}, then sufficiently far into the ultraviolet there is still a positive solution $\hat{g}_v$ to \eno{QPagain}, but as one proceeds toward the infrared, there comes a point where $1/\hat{f}_v$ gets so small that there is no real solution to \eno{QPagain}.  Thus it appears---at least from the analysis of the leading-order Schwinger-Dyson equation presented here---that the interacting theory is ill-defined for direction-dependent characters if we try to make the fields bosonic, even if we adjust $\sigma_\Omega$ to maintain the ultraviolet sign constraint \eno{UVsign}.  See figure~\ref{QuarticSoln}.

Most of the above discussion goes through with minor modifications when we consider direction-dependent characters on $\mathbb{Q}_2$.  In particular,
 \eqn{SFtwo}{
  \int_{\mathbb{Q}_2} dt \, \chi(\omega t) \delta_{v(t)} \sgn t = 
   M_\ell \delta_{v(\omega)-\ell} \sgn\omega \,,
 }
where $\ell = -2$ or $-3$ according to whether the integral for $\Gamma(\pi_{s,\sgn})$ localizes onto $2^{-2}\mathbb{U}_2$ or $2^{-3}\mathbb{U}_2$, and the constant $M_\ell$ is given by
\eqn{M-2}
{M_\ell\equiv 2^\ell\,\Gamma(\pi_{1,\sgn})=\pm \sqrt{2^\ell\sgn(-1)}.}
One easily sees that \eno{QuarticPolynomial} is modified to
 \eqn{QPtwo}{
  g_v = f_v + (\sigma_\psi 2^\ell g^2 N^3) p^{-2v} g_v^4 f_v \,.
 }
The discussion around \eno{IRgv} goes through as before, only with $p$ replaced by $2^{-\ell}$.  In particular, we must choose $\sigma_\psi = -1$ to get a sensible solution to the Schwinger-Dyson equation, and we find that $g_v$ has the same phase as $f_v$.

\section{A Wilsonian perspective}
\label{WILSON}

The Schwinger-Dyson equation \eno{SDGagain} for the two-point function lets us start with a theory whose microscopic description involves a pure power-law two-point function, $F(\omega) = \sqrt{\sgn(-1)}\,(\sgn\omega)/|\omega|^s$, and obtain from it a connected correlator $G(\omega)$ which interpolates between $F(\omega)$ in the ultraviolet and universal scaling $G(\omega) = b (\sgn\omega)/|\omega|^{1/2}$ in the infrared.  This is a crucial feature that drives much of the interest in the SYK model and its tensor model cousins: we flow from a free theory in the UV to a non-trivial IR fixed point with the emergent symmetries of $AdS_2$ (or more properly ``nearly" $AdS_2$). Over $\mathbb{R}$, these models are only soluble in the IR using the Archimedean version of the limiting-behavior analysis we performed in section~\ref{INFRARED}. For those non-Archimedean theories that are built with direction-dependent characters, we have shown in section~\ref{EXACT} that the Schwinger-Dyson equation has exact solutions interpolating between UV and IR behavior. 

However, there is an apparent paradox: We generically expect a non-renormalization theorem for bilocal kinetic terms.  So how is it that we are seeing $G(\omega)$ behave so differently from $F(\omega)$?  This apparent paradox is best examined in a Wilsonian context, where non-renormalization theorems have been cleanly demonstrated for $p$-adic field theories, e.g.~in \cite{Lerner:1989ty}.  In section~\ref{NONRENORMALIZATION} we will give some first indications that the expected non-renormalization theorem continues to hold in the presence of non-trivial sign characters.  For technical reasons to become apparent, we restrict our analysis in three ways.  First, we consider only the leading melonic limit.  Second, we require direction-dependent sign characters.  And third, we restrict $K$ to be $\mathbb{Q}_p$ or an extension of $\mathbb{Q}_p$ for odd $p$, or else $\mathbb{Q}_2$.  The upshot of our discussion is that the bilocal term in the Wilsonian effective action is indeed not renormalized, and as we explain in section~\ref{PARADOX} it is the distinction between this Wilsonian effective action and the one-particle irreducible (1PI) action that resolves the paradox.

\subsection{\padic non-renormalization}
\label{NONRENORMALIZATION}

A general feature of \padic field theories is that they behave more simply than their Archimedean counterparts under renormalization; indeed, they have strong commonalities with the hierarchical models studied in the early renormalization group literature. As explained by Lerner and Missarov \cite{Lerner:1989ty}, whose argument we will roughly follow in this subsection, there is a strong non-renormalization theorem which forbids non-trivial renormalization of the kinetic term.\footnote{A related point for Archimedean field theories is that bilocal kinetic terms like those of Fisher, Ma, and~Nickel \cite{Fisher:1972zz} don't get renormalized, and the reasoning in that case is that all divergences can be canceled by local counter terms, possibly using a derivative expansion.} This can be seen through the Wilsonian picture of perturbative renormalization in momentum space, where one attempts to define an effective field theory in the IR (external states having some low characteristic momentum $\omega$) by integrating out internal UV modes (loop momenta $\omega_i$ such that $\omega_i > \omega$). In particular, in evaluating loop corrections to the propagator (which typically lead to divergent integrals), we instead perform the integral over a shell of hard momenta of fixed magnitude $|\omega| = \Lambda$. This defines a new propagator, and iterating this procedure defines the renormalization group flow.

In \padic field theories, the ultrametric nature of the norm $|\omega|$ has a surprising and beneficial effect on the momentum shell renormalization technique. The ultrametric triangle inequality means that, when $\omega$ is soft and $\omega_i$ is hard,  $|\omega +\omega_i| = |\omega_i|$ exactly. This is in contrast to the weaker Archimedean  statement that $| \omega + \omega_i| \approx |\omega_i|$ when $\omega_i \gg \omega$.

The easiest way to see how this leads to non-renormalization is to work out the specific example relevant for this work: wavefunction renormalization in a melonic theory. The leading melonic correction to the quadratic part of the effective action is again the underground diagram of \eqref{LoopDiagram}. At the level of our current discussion, we only need the kinematic information, so we drop the flavor indices and use the following diagram:
 \eqn{RGUNDERGROUND}{
\begin{tikzpicture}[scale=1]
\newcommand{\midarrow}{\tikz \draw[-triangle 90] (0,0) -- +(.1,0);}
  \draw (2.7,-2) node[anchor=north] {} -- node[anchor = west]{} (-.7,-2) ;
\draw (-.7,-1.7) node[anchor=west] {$\omega$};
\draw (.85, -.7) node[anchor=west] {$\omega_1$};
\draw (.85,-1.7) node[anchor=west] {$\omega_2$};
\draw (.85,-2.7) node[anchor=west] {$\omega_3$};
\draw(1,-2) circle (1);
\draw(-.5, -2) node[anchor = west] {\midarrow};
\draw(.85, -2) node[anchor = west] {\midarrow};
\draw(.85, -1) node[anchor = west] {\midarrow};
\draw(.85, -3) node[anchor = west] {\midarrow};
\draw(2.2, -2) node[anchor = west] {\midarrow};
\end{tikzpicture}
}
Because we are correcting the effective action, we amputate the external propagators.  In a Wilsonian picture, we integrate out a shell of hard momenta $|\omega_i| = \Lambda$ and consider the soft external momentum $|\omega| < \Lambda$.  Through a rescaling of all momenta, we can choose $\Lambda = 1$.  Then, again dropping the $\Omega_{ab}$ factors, the amplitude from the diagram in \eno{RGUNDERGROUND} is proportional to
\begin{equation}
I_2(\omega) = \sigma_{\Omega}g^2 N^3\int_{{\cal O}_K^\times} \hspace{-2mm}  d \omega_1 d \omega_2 d \omega_3 \, \delta(\omega_1 + \omega_2 + \omega_3 - \omega) G(\omega_1) G(\omega_2) G(\omega_3)\,,
\label{I2omega}
\end{equation}
where ${\cal O}_K^\times$ is the multiplicative group of units in $K$.
If we are doing fixed-order perturbative renormalization to leading order in the melonic limit, then the $G(\omega_i)$ in \eno{I2omega} would be propagators of the free theory.  If instead we want resummed perturbation theory that includes all diagrams in the melonic limit, then the $G(\omega_i)$ would be an improved propagator of the cutoff theory that could be determined by a Schwinger-Dyson equation in this theory.  For the argument below to work, all that matters is that $G(\omega_i) = \hat{G}(|\omega_i|) \sgn\omega_i$ for some function $\hat{G}$ of the norm of $\omega_i$.

The key to the non-renormalization theorem is to make a $u$-substitution of the form
 \eqn{usub}{
  \tilde\omega_1 = \omega_1 - \omega \,,
 }
so as to arrive at
\begin{equation}
I_2(\omega) = \sigma_{\Omega}g^2 N^3\int_{\mathcal{O}_K^{\times}} \hspace{-2mm}  d \tilde\omega_1 d \omega_2 d \omega_3 \, \delta(\tilde\omega_1 + \omega_2 + \omega_3 ) G(\tilde\omega_1) G(\omega_2) G(\omega_3) = I_2(0)\,,
\label{I20}
\end{equation}
provided we can show $G(\omega_1) = G(\tilde\omega_1)$.  A key point is that the change of variables \eno{usub} is a measure-preserving bijection of ${\cal O}_K^\times$ to itself: this is where we use the ultrametric triangle inequality.

The absence of $\omega$ dependence from the last expression in \eno{I20} is the feature we seek: It means that the correction to the effective action has no $\omega$ dependence, and is instead a mass term for $\psi$, local in position space.  In other words, the bilocal term is not renormalized.  We may however feel some surprise at the appearance of a mass term in the effective action.  We will now argue---subject to aforementioned technical restrictions---that in fact $I_2(\omega) = 0$.

First let's handle the case where $K$ is (an extension of) $\mathbb{Q}_p$ for $p$ odd.  We write $\omega_i$ as in \eqref{tExpress}:
\eqn{wExpress}{
\omega_i = \mathfrak{p}^{v(\omega_i)} \epsilon^{w(\omega_i)} a(\omega_i)\,.
}
For $\omega_i \in \mathcal{O}_K^{\times}$, $v(\omega_i) = 0$, so the only quadratic characters available to us are the trivial character and $(-1)^{w(\omega_i)}$. Since we have the strict inequality $|\omega|<|\omega_1|$, shifting by $\omega$ cannot change the element of the residue field $\mathbb{F}_q$ and so $\text{sgn}(\tilde{\omega}_1) = \text{sgn}(\omega_1)$.  Because we assumed that $G(\omega_i) = \hat{G}(|\omega_i|) \sgn\omega_i$, this in turn implies $G(\tilde\omega_i) = G(\omega_i)$, and we see from \eno{I20} that $I_2(\omega)$ is independent of $\omega$.

To show that $I_2(\omega) = 0$, we need an additional argument that relies on the sign character in $G(\omega_i)$ being direction-dependent, i.e.~it must be $(-1)^{w(\omega_i)}$ rather than the trivial character.  Then there is some $\lambda \in {\cal O}_K^\times$ for which $\sgn\lambda = -1$.  Note that if $\omega_i$ covers $\Ok$ then so does $\lambda\omega_i$; to say it differently, $\lambda$ has a multiplicative inverse in~$\Ok$, so the  change of variables $\omega_i\to \lambda \omega_i$ is invertible. Performing it on~\eqref{I2omega}, we obtain
\eqn{eqn3}{
I_2(\omega) = (\sgn\lambda) I_2(\lambda^{-1} \omega) = -I_2(\omega) \,,
}
where we used the multiplicative property to simplify the sign characters.  
Since we already concluded that $I_2(\omega)$ is independent of $\omega$, \eno{eqn3} implies $I_2(\omega)=0$, as desired.

Next we consider the case $K=\mathbb{Q}_2$.  Because sign characters now depend on the second and third $2$-adic digits of their argument, it may not be true that $G(\omega_1) = G(\tilde\omega_1)$: Indeed, forming the difference $\omega_1-\omega$ may change the second or third digit of $\omega_1$, so this equality will fail for any direction-dependent character on $\mathbb{Q}_2$.  Nevertheless we can argue that $I_2(\omega) = 0$ for quite a trivial reason: it is impossible to have the three loop momenta $\omega_i \in {\cal O}_K^\times = \mathbb{U}_2$ add up to an external momentum $\omega$ with $|\omega| < 1$.  To see this, note that each $\omega_i = 1 + 2b_i$ for some $b_i \in \mathbb{Z}_2$, so also $\omega_1+\omega_2+\omega_3 = 1 + 2b$ for some $b \in \mathbb{Z}_2$.  In other words, the delta function \eno{I2omega} is always zero on the integration region we are using.  This argument shows that $I_2(\omega) = 0$ regardless of which character we use on $\mathbb{Q}_2$.

If $K$ is an extension of $\mathbb{Q}_2$, the argument is more difficult, and we do not know whether the bilocal term in the action is renormalized or not.  With the decomposition
 \eqn{ellDecomp}{
  \omega_i = \xi_i a_i \qquad\hbox{where\quad$\xi_i \in \mathbb{F}_K^\times$\quad and\quad$a_i \in A$\,,}
 }
we can re-express the $I_2(\omega)$ integral as
 \eqn{Itwo}{
  I_2(\omega) = \sigma_{\Omega}g^2 N^3 \sum_{\xi_i \in \mathbb{F}_K^\times} \delta_{\xi_1+\xi_2+\xi_3}
   \int_A da_1 da_2 da_3 \, \delta(\xi_1 a_1 + \xi_2 a_2 + \xi_3 a_3 - \omega)
     \prod_{i=1}^3 \sgn a_i \,.
 }
Here $\delta_\xi$ is the Kronecker delta function, which arises because $\xi_1 a_1 + \xi_2 a_2 + \xi_3 a_3 = \omega$ implies $\xi_1+\xi_2+\xi_3 = 0$ when $|\omega|<1$.  If $\mathbb{F}_K = \mathbb{F}_2$, as in $\mathbb{Q}_2$ or its totally ramified extensions, then $\delta_{\xi_1+\xi_2+\xi_3} = 0$ always, and we once again have $I_2(\omega) = 0$.  In the general case, where $\mathbb{F}_K = \mathbb{F}_{2^f}$ for some $f>1$, it is not clear to us whether $I_2(\omega)$ vanishes.

\subsection{Resolution of an apparent paradox}
\label{PARADOX}

So far, we have shown in the leading melonic limit that the correction $I_2(\omega)$ to the Wilsonian effective action vanishes, at least for a broad range of choices of field $K$ and sign character.  
This seems at first puzzling when we try to interpret the renormalization group equations, which in the \padic setting are a discrete set of recursion relations that relate the effective action with cutoff $\Lambda$ to the effective action with cutoff $\Lambda / p$. Iterating these equations defines the renormalization group flow. With $I_2(\omega) = 0$ in our models, the quadratic part of the effective action isn't any different from what we started with, so there appears to be no flow.
However,
as we have already seen explicitly in section~\ref{EXACT} for direction-dependent characters, the Schwinger-Dyson equation has a solution $G$ interpolating between the free UV theory and the universal IR scaling behavior.  This is the apparent paradox that we referred to before.

The resolution is the distinction between the 1PI effective action and the Wilsonian effective action.  The kernel of the quadratic part of the Wilsonian effective action is not renormalized (subject to the technical assumptions described previously), but added to it is a relevant $g \psi^4$ interaction term which becomes large in the infrared.\footnote{In the leading melonic limit, this quartic term is also not renormalized: for example, the first correction is order $g^2 N = {g \over \sqrt{N}} (g^2 N^3)^{1/2}$, which is indeed subleading.  This non-renormalization property doesn't matter particularly to our story; what matters is that the interaction always remains relevant so that its effects in the infrared are large.}  For this reason, in the infrared, the quadratic kernel of the Wilsonian effective action strongly disagrees with the kernel of the quantum effective action. In this limit we are far from a Gaussian theory that would be defined by naive RG flow. The solution of the Schwinger-Dyson equation \eno{SDGagain} shows that this kind of theory can have interesting dependence on scale beyond the conventional wisdom of effective field theory; and this is particularly striking for ultrametric theories.

\section{Summary and future directions}
\label{SUMMARY}

Melonic theories are simple because the list of diagrams we must compute in the leading order melonic limit is very short: For the two-point function, all of them can be resummed into the Schwinger-Dyson equation \eno{SDGagain}.  Defining melonic theories over the $p$-adics makes the theories in some cases even simpler, because one can then solve the Schwinger-Dyson equation exactly to find the interpolation between the free theory in the ultraviolet and universal infrared power law scaling in the infrared.  As explained in section~\ref{EXACT}, the key to this exact solution is to choose a sign function $\sgn t$ on $\mathbb{Q}_p$ so that there exists some number $\lambda$ with $|\lambda|=1$ and $\sgn\lambda = -1$.  We previously described these sign functions as direction-dependent.

A benefit of formulating a melonic version of the Klebanov-Tarnopolsky model on the $p$-adics is that it naturally leads to a cleaner and more unified description of the Archimedean theory, as well as a generalization to theories with bilocal kinetic terms.  Indeed, our description in \eno{KTfull}-\eno{Sint} of the theories we consider simultaneously encompasses bosonic and fermionic theories on $\mathbb{R}$, the bosonic theory on $\mathbb{C}$, and bosonic and fermionic theories on all the $\mathbb{Q}_p$ and their extensions.  Moreover, the normalization constant $b$ in the infrared limit of the two-point function, $G(t) = b {\sgn t \over |t|^{1/2}}$, is seen to be determined in terms of the product $\Gamma(\pi_{-{1 \over 2},\sgn}) \Gamma(\pi_{{1 \over 2},\sgn})$ of two generalized gamma functions, as indicated in \eno{PidSolve}.  Strikingly, $b$ does not depend on the spectral index of the ultraviolet theory.

For $\mathbb{R}$, $\mathbb{C}$, and $\mathbb{Q}_p$, our understanding of generalized gamma functions is sufficiently complete that we can determine the sign of the crucial quantity $\Gamma(\pi_{-{1 \over 2},\sgn}) \Gamma(\pi_{{1 \over 2},\sgn})$ in all cases.  This sign, together with reality properties of the two-point function, forces us to choose the statistics of the field $\psi^{abc}$ to be bosonic if $\sgn$ is the trivial sign function (setting the sign of all non-zero numbers to $1$), and fermionic otherwise.  Related constraints force us to modify the ordinary ${\rm O}(N)^3$ symmetry of melonic theories to ${\rm Sp}(N)^3$, where ${\rm Sp}(N)$ is the non-compact group of $N \times N$ matrices preserving a symplectic form.  Altogether, we manage to construct precisely one type of melonic theory in every place ($\mathbb{R}$ or $\mathbb{Q}_p$ for any $p$) and for every choice of sign function.  All of them have the same characteristic scaling of the two-point function in the infrared, $G(t) = b {\sgn t \over |t|^{1/2}}$ in the limit of large $|t|$.

An obvious generalization of our results would be to pass to arbitrary extensions of $\mathbb{Q}_p$, where our current understanding of generalized gamma functions is insufficient to determine even the phase of $\Gamma(\pi_{-{1 \over 2},\sgn}) \Gamma(\pi_{{1 \over 2},\sgn})$.  In order to wind up with bosonic theories when the sign function is trivial and fermionic theories otherwise, we would need $\Gamma(\pi_{-{1 \over 2},\sgn}) \Gamma(\pi_{{1 \over 2},\sgn})$ always to be real, and we would need its sign to be $-1$ for the trivial sign function and $\sgn(-1)$ for non-trivial sign functions.  It would be interesting to find a way to prove that the signs of $\Gamma(\pi_{-{1 \over 2},\sgn}) \Gamma(\pi_{{1 \over 2},\sgn})$ are indeed as we have just described.

Just as it is useful to study the SYK model and its relatives on $S^1$, another obvious future direction is to study our models on quotients of $\mathbb{Q}_p$.  This is likely to be quite a rich subject because there are many such quotients: Dividing $\mathbb{Q}_p$ (or more properly $\mathbb{P}^1(\mathbb{Q}_p)$ by an appropriate discrete subgroup of ${\rm PGL}(2,\mathbb{Q}_p)$ leads to a Mumford curve over $\mathbb{Q}_p$, and correlators on such spaces are likely to be related to automorphic functions.

Just as our results on the Schwinger-Dyson equation admit some obvious generalizations, our results on the Wilsonian renormalization of $p$-adic melonic theories leave considerable room for further work.  We showed that the bilocal kinetic term is not renormalized in the leading melonic limit in an assortment of theories with direction-dependent sign characters; however, we never progressed to more complicated graphs.  Existing treatments, e.g.~\cite{Lerner:1989ty}, go much further for theories that do not involve sign characters, demonstrating all-orders perturbative non-renormalization theorems.  It is tempting to think that similar results could be demonstrated in the theories we consider; however, particularly for extensions of $\mathbb{Q}_2$, the arguments seem non-trivial.  A full account of renormalization group properties of melonic theories should also include the possibility of a mass term, which we have suppressed in our treatment in order to obtain power-law scaling in the infrared, but which is sometimes permitted by the symmetries.

We end by noting that the space of theories we consider has non-trivial topology when the field $K$ is ultrametric.  The non-compact direction is parametrized by the spectral parameter $s$, which can be understood as controlling the ultraviolet limit of the two-point function: $G(t) \sim {\sgn t \over |t|^{1-s}}$ for small $|t|$, and we require $s>1/2$ in order for the $g \psi^4$ interaction to be relevant, while $s \leq 1$ in order to avoid a negative dimension for $\psi$.  For simplicity let's also assume $g>0$.  If $K=\mathbb{R}$ or $\mathbb{C}$, then we can rescale $g$ to $1$ by applying a scale transformation to the whole theory.  If $K$ is ultrametric, then the allowable scale transformations are discrete, consisting of multiplying all frequencies $\omega$ by an integer power of the uniformizer $\mathfrak{p}$.  The simplest such transformation is $\omega \to \mathfrak{p} \omega$, and all real quantities must at the same time be rescaled as $X \to p^{-[X]} X$.  For example, $g \to p^{-2s+1} g$.  Thus $g=p^{2s-1}$ is equivalent to $g=1$, but theories with $g \in (1,p^{2s-1})$ are all inequivalent.  In short, the space of values of $g$ is really a circle, constructed by identifying the endpoints of the interval $[1,p^{2s-1}]$.  Overall, the space of ultraviolet theories with $g>0$ has the topology of a cylinder.  It would be interesting to understand whether this topology has any significance in a non-perturbative treatment.

\subsection*{Acknowledgments}

We thank M.~Marcolli and P.~Witaszczyk for extensive discussions. B.S.\ would also like to thank A.~Almheiri for useful discussions. The work of S.S.G., C.J., S.P.\ and B.T.\ was supported in part by the Department of Energy under Grant No.~DE-FG02-91ER40671. The work of S.P.\ was also supported in part by the Bershadsky Family Fellowship Fund in Mathematics or Physics. The work of M.H.\ was supported by the U.S. Department of Energy, Office of Science, Office of High Energy Physics, under Award Number DE-SC0011632 as well as by the Walter Burke Institute for Theoretical Physics at Caltech. The work of B.S.\ was supported in part by the Simons Foundation, and by the U.S. Department of Energy under grant DE-SC-0009987. B.S.\ would like to thank the Stanford Institute for Theoretical Physics at Stanford University and the Aspen Center for Physics for hospitality. The work of B.S.\ was performed in part at Aspen Center for Physics, which is supported by National Science Foundation grant PHY-1607611.

\clearpage
\appendix
\section{Conventions for ${\rm O}(N)$ and ${\rm Sp}(N)$}
\label{SYMPLECTIC}

Throughout this text we make use of an invariant tensor $\Omega_{ab}$ of ${\rm GL}(N, \mathbb{R})$ to contract flavor indices of the fields. Usually these are Majorana fermions transforming in the trifundamental representation of the flavor group:
\begin{equation}
\psi^{abc} \rightarrow (\Lambda_1)^a{}_{a'} (\Lambda_2)^b{}_{b'} (\Lambda_3)^c{}_{c'} \psi^{a' b' c'}.
\end{equation}
where $\Lambda^a{}_b$ are the group matrices preserving this structure $\Omega_{ab} = \Lambda^c{}_a \Lambda^d{}_b \Omega_{cd}$. In order to preserve the reality condition on the fermions, we use real valued matrices, and this restricts which kinds of flavor groups can appear.

Interesting choices for $\Omega$ are those of definite parity. The symmetric choice is defined by $\Omega = \Omega^T$ and has $\sigma_\Omega = 1$.  The condition imposed on the group matrices defines the orthogonal group ${\rm O}(N)$. 

For the case of antisymmetric $\Omega = - \Omega^T$ with $\sigma_\Omega = -1$, we must take $N$ to be even and define
\begin{equation}
{\rm Sp}(N) \equiv \left \{ \Lambda \in {\rm SL}(N,\mathbb{R}): \Lambda^{T} \Omega \Lambda = \Omega \right \}.
\end{equation}
Elsewhere in the literature, this real symplectic group is sometimes denoted ${\rm Sp}(2N, \mathbb{R})$ to emphasize the matrices are real valued and even dimensional. This group is connected, but \emph{non-compact}.

There is a different but closely related kind of symplectic group called ${\rm USp}(2N)$ (also called ${\rm Sp}(N)$ in some references!) In contrast to the symplectic group above, ${\rm USp}(2N)$ is defined by the intersection
\begin{equation}
{\rm USp}(2N) = {\rm U}(2N) \cap {\rm Sp}(2N, \mathbb{C}).
\end{equation}
This is a compact group, but as complex matrices they cannot be used for the transformation of Majorana fermions.

In SYK-like models, only singlet states of the fermions (e.g. certain bilinears) have bulk duals and the fundamental fermions themselves don't seem to have any bulk interpretation. One advantage of tensor models is that we can understand the restriction to singlets by gauging the flavor group \cite{Witten:2016iux} \cite{Klebanov:2016xxf}. In $0+1$ dimensions the gauge field has no dynamics and appears only as an auxiliary field. Integrating it out projects onto the singlet states. 

In the $p$-adic case, the non-Archimedean structure of $\mathbb{Q}_p$ makes conserved currents somewhat tricky to define. Additionally, one might worry about gauging a non-compact group such as ${\rm Sp}(N)$ which would have problematic ghosts in higher dimensions. For the purposes of this work we ignore the issues of gauging and treat the ${\rm O}(N)^3$ and ${\rm Sp}(N)^3$ symmetries as global flavor symmetries.

\bibliographystyle{ssg}
\bibliography{signed} 
\end{document}